\def\figflag{y} \input epsf
\def\boringfonts{y}  

%
%
\def\unredoffs{} \def\redoffs{\voffset=-.31truein\hoffset=-13truemm}
\def\speclscape{}
%
%
%
%
%
\newbox\leftpage \newdimen\fullhsize \newdimen\hstitle \newdimen\hsbody
\tolerance=1000\hfuzz=2pt
\catcode`\@=11 
\def\bigans{b }
\message{ big or little (b/l)? }\read-1 to\answ
\ifx\answ\bigans\message{(This will come out unreduced.}
\magnification=1200\unredoffs\baselineskip=16pt plus 2pt minus 1pt
\hsbody=\hsize \hstitle=\hsize 
\else\message{(This will be reduced.} \let\l@r=L
\magnification=1000\baselineskip=16pt plus 2pt minus 1pt \vsize=7truein
\redoffs \hstitle=8truein\hsbody=4.75truein\fullhsize=10truein\hsize=\hsbody
\output={\ifnum\pageno=0 
  \shipout\vbox{\speclscape{\hsize\fullhsize\makeheadline}
    \hbox to \fullhsize{\hfill\pagebody\hfill}}\advancepageno
  \else
  \almostshipout{\leftline{\vbox{\pagebody\makefootline}}}\advancepageno 
  \fi}
\def\almostshipout#1{\if L\l@r \count1=1 \message{[\the\count0.\the\count1]}
      \global\setbox\leftpage=#1 \global\let\l@r=R
 \else \count1=2
  \shipout\vbox{\speclscape{\hsize\fullhsize\makeheadline}
      \hbox to\fullhsize{\box\leftpage\hfil#1}}  \global\let\l@r=L\fi}
\fi
%
\newcount\yearltd\yearltd=\year\advance\yearltd by -1900

\def\Title#1#2{\nopagenumbers\abstractfont\hsize=\hstitle\rightline{#1}%
\vskip 1in\centerline{\titlefont #2}\abstractfont\vskip .5in\pageno=0}
\def\Date#1{\vfill\leftline{#1}\tenpoint\supereject\global\hsize=\hsbody%
\footline={\hss\tenrm\folio\hss}}
%

\def\draftmode{\message{ DRAFTMODE }\def\draftdate{{\rm preliminary draft:
\number\month/\number\day/\number\yearltd\ \ \hourmin}}%
\headline={\hfil\draftdate}\writelabels\baselineskip=20pt plus 2pt minus 2pt
 {\count255=\time\divide\count255 by 60 \xdef\hourmin{\number\count255}
  \multiply\count255 by-60\advance\count255 by\time
  \xdef\hourmin{\hourmin:\ifnum\count255<10 0\fi\the\count255}}}
\def\nolabels{\def\wrlabeL##1{}\def\eqlabeL##1{}\def\reflabeL##1{}}
\def\writelabels{\def\wrlabeL##1{\leavevmode\vadjust{\rlap{\smash%
{\line{{\escapechar=` \hfill\rlap{\sevenrm\hskip.03in\string##1}}}}}}}%
\def\eqlabeL##1{{\escapechar-1\rlap{\sevenrm\hskip.05in\string##1}}}%
\def\reflabeL##1{\noexpand\llap{\noexpand\sevenrm\string\string\string##1}}}
\nolabels
%
\global\newcount\secno \global\secno=0
\global\newcount\meqno \global\meqno=1
\def\newsec#1{\global\advance\secno by1\message{(\the\secno. #1)}
\global\subsecno=0\eqnres@t\noindent{\bf\the\secno. #1}
\writetoca{{\secsym} {#1}}\par\nobreak\medskip\nobreak}
\def\eqnres@t{\xdef\secsym{\the\secno.}\global\meqno=1\bigbreak\bigskip}
\def\sequentialequations{\def\eqnres@t{\bigbreak}}\xdef\secsym{}
\global\newcount\subsecno \global\subsecno=0
\def\subsec#1{\global\advance\subsecno by1\message{(\secsym\the\subsecno. #1)}
\ifnum\lastpenalty>9000\else\bigbreak\fi
\noindent{\it\secsym\the\subsecno. #1}\writetoca{\string\quad 
{\secsym\the\subsecno.} {#1}}\par\nobreak\medskip\nobreak}
\def\appendix#1#2{\global\meqno=1\global\subsecno=0\xdef\secsym{\hbox{#1.}}
\bigbreak\bigskip\noindent{\bf Appendix #1. #2}\message{(#1. #2)}
\writetoca{Appendix {#1.} {#2}}\par\nobreak\medskip\nobreak}
%
%
\def\eqnn#1{\xdef #1{(\secsym\the\meqno)}\writedef{#1\leftbracket#1}%
\global\advance\meqno by1\wrlabeL#1}
\def\eqna#1{\xdef #1##1{\hbox{$(\secsym\the\meqno##1)$}}
\writedef{#1\numbersign1\leftbracket#1{\numbersign1}}%
\global\advance\meqno by1\wrlabeL{#1$\{\}$}}
\def\eqn#1#2{\xdef #1{(\secsym\the\meqno)}\writedef{#1\leftbracket#1}%
\global\advance\meqno by1$$#2\eqno#1\eqlabeL#1$$}
%
\newskip\footskip\footskip14pt plus 1pt minus 1pt 
\def\footnotefont{\ninepoint}\def\f@t#1{\footnotefont #1\@foot}
\def\f@@t{\baselineskip\footskip\bgroup\footnotefont\aftergroup\@foot\let\next}
\setbox\strutbox=\hbox{\vrule height9.5pt depth4.5pt width0pt}
\global\newcount\ftno \global\ftno=0
\def\foot{\global\advance\ftno by1\footnote{$^{\the\ftno}$}}
%
\newwrite\ftfile   
\def\footend{\def\foot{\global\advance\ftno by1\chardef\wfile=\ftfile
$^{\the\ftno}$\ifnum\ftno=1\immediate\openout\ftfile=foots.tmp\fi%
\immediate\write\ftfile{\noexpand\smallskip%
\noexpand\item{f\the\ftno:\ }\pctsign}\findarg}%
\def\footatend{\vfill\eject\immediate\closeout\ftfile{\parindent=20pt
\centerline{\bf Footnotes}\nobreak\bigskip\input foots.tmp }}}
\def\footatend{}
%
%
\global\newcount\refno \global\refno=1
\newwrite\rfile
\def\ref{[\the\refno]\nref}
\def\nref#1{\xdef#1{[\the\refno]}\writedef{#1\leftbracket#1}%
\ifnum\refno=1\immediate\openout\rfile=refs.tmp\fi
\global\advance\refno by1\chardef\wfile=\rfile\immediate
\write\rfile{\noexpand\item{#1\ }\reflabeL{#1\hskip.31in}\pctsign}\findarg}
\def\findarg#1#{\begingroup\obeylines\newlinechar=`\^^M\pass@rg}
{\obeylines\gdef\pass@rg#1{\writ@line\relax #1^^M\hbox{}^^M}%
\gdef\writ@line#1^^M{\expandafter\toks0\expandafter{\striprel@x #1}%
\edef\next{\the\toks0}\ifx\next\em@rk\let\next=\endgroup\else\ifx\next\empty%
\else\immediate\write\wfile{\the\toks0}\fi\let\next=\writ@line\fi\next\relax}}
\def\striprel@x#1{} \def\em@rk{\hbox{}} 
\def\lref{\begingroup\obeylines\lr@f}
\def\lr@f#1#2{\gdef#1{\ref#1{#2}}\endgroup\unskip}

\def\addref#1{\immediate\write\rfile{\noexpand\item{}#1}} 
\def\startrefs#1{\immediate\openout\rfile=refs.tmp\refno=#1}
\def\xref{\expandafter\xr@f}\def\xr@f[#1]{#1}
\def\refs#1{\count255=1[\r@fs #1{\hbox{}}]}
\def\r@fs#1{\ifx\und@fined#1\message{reflabel \string#1 is undefined.}%
\nref#1{need to supply reference \string#1.}\fi%
\vphantom{\hphantom{#1}}\edef\next{#1}\ifx\next\em@rk\def\next{}%
\else\ifx\next#1\ifodd\count255\relax\xref#1\count255=0\fi%
\else#1\count255=1\fi\let\next=\r@fs\fi\next}
%

%
\newwrite\ffile\global\newcount\figno \global\figno=1
\def\fig{fig.~\the\figno\nfig}
\def\nfig#1{\xdef#1{fig.~\the\figno}%
\writedef{#1\leftbracket fig.\noexpand~\the\figno}%
\ifnum\figno=1\immediate\openout\ffile=figs.tmp\fi\chardef\wfile=\ffile%
\immediate\write\ffile{\noexpand\medskip\noexpand\item{Fig.\ \the\figno. }
\reflabeL{#1\hskip.55in}\pctsign}\global\advance\figno by1\findarg}
\def\xfig{\expandafter\xf@g}\def\xf@g fig.\penalty\@M\ {}
\def\figs#1{figs.~\f@gs #1{\hbox{}}}
\def\f@gs#1{\edef\next{#1}\ifx\next\em@rk\def\next{}\else
\ifx\next#1\xfig #1\else#1\fi\let\next=\f@gs\fi\next}
\newwrite\lfile
{\escapechar-1\xdef\pctsign{\string\%}\xdef\leftbracket{\string\{}
\xdef\rightbracket{\string\}}\xdef\numbersign{\string\#}}

\def\writestop{\def\writestoppt{\immediate\write\lfile{\string\pageno%
\the\pageno\string\startrefs\leftbracket\the\refno\rightbracket%
\string\def\string\secsym\leftbracket\secsym\rightbracket%
\string\secno\the\secno\string\meqno\the\meqno}\immediate\closeout\lfile}}
\def\writestoppt{}\def\writedef#1{}
\def\seclab#1{\xdef #1{\the\secno}\writedef{#1\leftbracket#1}\wrlabeL{#1=#1}}
\def\subseclab#1{\xdef #1{\secsym\the\subsecno}%
\writedef{#1\leftbracket#1}\wrlabeL{#1=#1}}
\newwrite\tfile \def\writetoca#1{}
\def\leaderfill{\leaders\hbox to 1em{\hss.\hss}\hfill}
\def\writetoc{\immediate\openout\tfile=toc.tmp 
   \def\writetoca##1{{\edef\next{\write\tfile{\noindent ##1 
   \string\leaderfill {\noexpand\number\pageno} \par}}\next}}}

%
\catcode`\@=12 
%
\edef\tfontsize{\ifx\answ\bigans scaled\magstep3\else scaled\magstep4\fi}
\font\titlerm=cmr10 \tfontsize \font\titlerms=cmr7 \tfontsize
\font\titlermss=cmr5 \tfontsize \font\titlei=cmmi10 \tfontsize
\font\titleis=cmmi7 \tfontsize \font\titleiss=cmmi5 \tfontsize
\font\titlesy=cmsy10 \tfontsize \font\titlesys=cmsy7 \tfontsize
\font\titlesyss=cmsy5 \tfontsize \font\titleit=cmti10 \tfontsize
\skewchar\titlei='177 \skewchar\titleis='177 \skewchar\titleiss='177
\skewchar\titlesy='60 \skewchar\titlesys='60 \skewchar\titlesyss='60
\def\titlefont{\def\rm{\fam0\titlerm}
\textfont0=\titlerm \scriptfont0=\titlerms \scriptscriptfont0=\titlermss
\textfont1=\titlei \scriptfont1=\titleis \scriptscriptfont1=\titleiss
\textfont2=\titlesy \scriptfont2=\titlesys \scriptscriptfont2=\titlesyss
\textfont\itfam=\titleit \def\it{\fam\itfam\titleit}\rm}
 \ifx\answ\bigans\else scaled\magstep1\fi
\ifx\answ\bigans\def\abstractfont{\tenpoint}\else
\font\abssl=cmsl10 scaled \magstep1
\font\absrm=cmr10 scaled\magstep1 \font\absrms=cmr7 scaled\magstep1
\font\absrmss=cmr5 scaled\magstep1 \font\absi=cmmi10 scaled\magstep1
\font\absis=cmmi7 scaled\magstep1 \font\absiss=cmmi5 scaled\magstep1
\font\abssy=cmsy10 scaled\magstep1 \font\abssys=cmsy7 scaled\magstep1
\font\abssyss=cmsy5 scaled\magstep1 \font\absbf=cmbx10 scaled\magstep1
\skewchar\absi='177 \skewchar\absis='177 \skewchar\absiss='177
\skewchar\abssy='60 \skewchar\abssys='60 \skewchar\abssyss='60
\def\abstractfont{\def\rm{\fam0\absrm}
\textfont0=\absrm \scriptfont0=\absrms \scriptscriptfont0=\absrmss
\textfont1=\absi \scriptfont1=\absis \scriptscriptfont1=\absiss
\textfont2=\abssy \scriptfont2=\abssys \scriptscriptfont2=\abssyss
\textfont\itfam=\bigit \def\it{\fam\itfam\bigit}\def\footnotefont{\tenpoint}%
\textfont\slfam=\abssl \def\sl{\fam\slfam\abssl}%
\textfont\bffam=\absbf \def\bf{\fam\bffam\absbf}\rm}\fi
\def\tenpoint{\def\rm{\fam0\tenrm}
\textfont0=\tenrm \scriptfont0=\sevenrm \scriptscriptfont0=\fiverm
\textfont1=\teni  \scriptfont1=\seveni  \scriptscriptfont1=\fivei
\textfont2=\tensy \scriptfont2=\sevensy \scriptscriptfont2=\fivesy
\textfont\itfam=\tenit \def\it{\fam\itfam\tenit}\def\footnotefont{\ninepoint}%
\textfont\bffam=\tenbf \def\bf{\fam\bffam\tenbf}\def\sl{\fam\slfam\tensl}\rm}
\font\ninerm=cmr9  \font\ninei=cmmi9 \font\sixi=cmmi6 
\font\ninesy=cmsy9 \font\sixsy=cmsy6 \font\ninebf=cmbx9 
\font\nineit=cmti9 \font\ninesl=cmsl9 \skewchar\ninei='177
\skewchar\sixi='177 \skewchar\ninesy='60 \skewchar\sixsy='60 
\def\ninepoint{\def\rm{\fam0\ninerm}
\textfont0=\ninerm \scriptfont0=\sixrm \scriptscriptfont0=\fiverm
\textfont1=\ninei \scriptfont1=\sixi \scriptscriptfont1=\fivei
\textfont2=\ninesy \scriptfont2=\sixsy \scriptscriptfont2=\fivesy
\textfont\itfam=\ninei \def\it{\fam\itfam\nineit}\def\sl{\fam\slfam\ninesl}%
\textfont\bffam=\ninebf \def\bf{\fam\bffam\ninebf}\rm} 
%
%
\def\noblackbox{\overfullrule=0pt}
\hyphenation{anom-aly anom-alies coun-ter-term coun-ter-terms}
\def\inv{^{\raise.15ex\hbox{${\scriptscriptstyle -}$}\kern-.05em 1}}

\def\Dsl{\,\raise.15ex\hbox{/}\mkern-13.5mu D} 
\def\dsl{\raise.15ex\hbox{/}\kern-.57em\partial}
\def\del{\partial}

\catcode`\@=11 

\def\slashit#1{\mathord{\mathpalette\c@ncel{#1}}}
\catcode`\@=12
\def\tr{{\rm tr}} \def\Tr{{\rm Tr}}
\font\bigit=cmti10 scaled \magstep1
\def\lspace{\ifx\answ\bigans{}\else\qquad\fi}
\def\lbspace{\ifx\answ\bigans{}\else\hskip-.2in\fi} 
\def\boxeqn#1{\vcenter{\vbox{\hrule\hbox{\vrule\kern3pt\vbox{\kern3pt
	\hbox{${\displaystyle #1}$}\kern3pt}\kern3pt\vrule}\hrule}}}
\def\mbox#1#2{\vcenter{\hrule \hbox{\vrule height#2in
		\kern#1in \vrule} \hrule}}  
%
 \def\CO{{\cal O}} 
    
 \def\CH{{\cal H}}

\def\darr#1{\raise1.5ex\hbox{$\leftrightarrow$}\mkern-16.5mu #1}

\def\half{{\textstyle{1\over2}}} 
\def\roughly#1{\raise.3ex\hbox{$#1$\kern-.75em\lower1ex\hbox{$\sim$}}}


\def\fonttest{y}
\ifx\boringfonts\fonttest\else
\font\ninerm=cmr9\font\ninei=cmmi9\font\nineit=cmti9\font\ninesy=cmsy9
\font\ninebf=cmbx9\font\ninesl=cmsl9
\def\ninepoint{\def\rm{\fam0\ninerm}
\textfont0=\ninerm \scriptfont0=\sevenrm \scriptscriptfont0=\fiverm
\textfont1=\ninei  \scriptfont1=\seveni  \scriptscriptfont1=\fivei
\textfont2=\ninesy \scriptfont2=\sevensy \scriptscriptfont2=\fivesy
\textfont\itfam=\nineit \def\it{\fam\itfam\nineit} \def\sl{\fam\slfam\ninesl}
\textfont\bffam=\ninebf \def\bf{\fam\bffam\ninebf} \rm}
\fi

\hyphenation{anom-aly anom-alies coun-ter-term coun-ter-terms
dif-feo-mor-phism dif-fer-en-tial super-dif-fer-en-tial dif-fer-en-tials
super-dif-fer-en-tials reparam-etrize param-etrize reparam-etriza-tion}
 
 
%
%
%
\newwrite\tocfile\global\newcount\tocno\global\tocno=1
\ifx\bigans\answ \def\tocline#1{\hbox to 320pt{\hbox to 45pt{}#1}}
\else\def\tocline#1{\line{#1}}\fi
\def\toclead{\leaders\hbox to 1em{\hss.\hss}\hfill}
\def\tnewsec#1#2{\xdef #1{\the\secno}\newsec{#2}
\ifnum\tocno=1\immediate\openout\tocfile=toc.tmp\fi\global\advance\tocno
by1
{\let\the=0\edef\next{\write\tocfile{\medskip\tocline{\secsym\ #2\toclead\the
\count0}\smallskip}}\next}
}
\def\tnewsubsec#1#2{\xdef #1{\the\secno.\the\subsecno}\subsec{#2}
\ifnum\tocno=1\immediate\openout\tocfile=toc.tmp\fi\global\advance\tocno
by1
{\let\the=0\edef\next{\write\tocfile{\tocline{ \ \secsym\the\subsecno\
#2\toclead\the\count0}}}\next}
}
\def\tappendix#1#2#3{\xdef #1{#2.}\appendix{#2}{#3}
\ifnum\tocno=1\immediate\openout\tocfile=toc.tmp\fi\global\advance\tocno
by1
{\let\the=0\edef\next{\write\tocfile{\tocline{ \ #2.
#3\toclead\the\count0}}}\next}
}
%
%

%
%
%
%

%
\long\def\optional#1{}
\def\cmp#1{#1}         
%
%
\let\narrowequiv=\equiv
\def\equiv{\;\narrowequiv\;}

\def\tilde{\widetilde}
\fontdimen16\tensy=2.7pt\fontdimen17\tensy=2.7pt 
 
 
 
%
\def\ga{\gamma}
\def\la{\lambda}
\def\ep{\epsilon}
\def\al{{\alpha}}

\def\dl{\delta}
%
%

\def\CJ{{\cal J}}\def\CCJ{{$\cal J$}}

\def\CO{{\cal O}}
\def\Order{{\oldcal O}}

\def\CH{{\cal H}}

%
%
%
\def\boxit#1#2{
        $$\vcenter{\vbox{\hrule\hbox{\vrule\kern3pt\vbox{\kern3pt
	\hbox to #1truein{\hsize=#1truein\vbox{#2}}\kern3pt}\kern3pt\vrule}
        \hrule}}$$
}


%

\def\lfr#1#2{{\textstyle{#1\over#2}}} 



\def\splitexact#1#2{\mathrel{\mathop{\null{
\lower4pt\hbox{$\rightarrow$}\atop\raise4pt\hbox{$\leftarrow$}}}\limits
^{#1}_{#2}}}

%
%
\def\pa{\partial}

\def\pd#1#2{{\partial #1\over\partial #2}} 
%
%
\def\ub{{\bar{\vphantom\i u}}}  
%
%

\def\ex#1{{\rm e}^{#1}}                 
\def\dd{\mskip 1.3mu{\rm d}\mskip .7mu} 
\def\tr{\hbox{tr}}                      


\def\det{\hbox{det$\,$}}                

%
%

\def\IM{isomorphism}

\def\eg{{\it e.g.}}\def\ie{{\it i.e.}}\def\etc{{\it etc.}}

%
%

\ifx\boringfonts\fonttest
\font\blackboard=cmssbx10 \font\blackboards=cmssbx10 at 7pt  
\font\blackboardss=cmssbx10 at 5pt
\else 
\font\blackboard=msym10 \font\blackboards=msym7   
\font\blackboardss=msym5
\fi
\newfam\black
\textfont\black=\blackboard
\scriptfont\black=\blackboards
\scriptscriptfont\black=\blackboardss

\let\spec=\blackb                                  
\ifx\boringfonts\fonttest
\font\gothic=cmssbx10 \font\gothics=cmssbx10 at 7pt  
\font\gothicss=cmssbx10 at 5pt
\else
\font\gothic=eufm10 \font\gothics=eufm7
\font\gothicss=eufm5
\fi
\newfam\gothi
\textfont\gothi=\gothic
\scriptfont\gothi=\gothics
\scriptscriptfont\gothi=\gothicss

{\catcode`\@=11\gdef\oldcal{\fam\tw@}}
\newfam\curly
\ifx\boringfonts\fonttest\else
\font\curlyfont=eusm10 \font\curlyfonts=eusm7
\font\curlyfontss=eusm5
\textfont\curly=\curlyfont
\scriptfont\curly=\curlyfonts
\scriptscriptfont\curly=\curlyfontss
\def\cal{\fam\curly\relax}    
\fi
%

\ifx\boringfonts\fonttest\else\fi

\global\newcount\pnfigno \global\pnfigno=1
\newwrite\ffile
\def\pfig#1#2{Fig.~\the\pnfigno\pnfig#1{#2}}
\def\pnfig#1#2{\xdef#1{Fig. \the\pnfigno}%
\ifnum\pnfigno=1\immediate\openout\ffile=figs.tmp\fi%
\immediate\write\ffile{\noexpand\item{\noexpand#1\ }#2}%
\global\advance\pnfigno by1}

%
%
\def\tfig#1{Fig.~\the\pnfigno\xdef#1{Fig.~\the\pnfigno}\global\advance\pnfigno by1}

%
%
%
%
\def\figI{y}
\def\ifigure#1#2#3#4{
\midinsert
\ifx\figflag\figI
\vbox to #4truein{
\vfil\centerline{\epsfysize=#4truein\epsfbox{#3}}}
\else
\vbox to .2truein{}
\fi
\narrower\narrower\noindent{\bf #1:} #2
\endinsert
}








%
%

%


\def\inbar{\,\vrule height1.5ex width.4pt depth0pt}
\def\IB{\relax{\rm I\kern-.18em B}}
\def\IC{\relax\hbox{$\inbar\kern-.3em{\rm C}$}}
\def\ID{\relax{\rm I\kern-.18em D}}
\def\IE{\relax{\rm I\kern-.18em E}}
\def\IF{\relax{\rm I\kern-.18em F}}
\def\IG{\relax\hbox{$\inbar\kern-.3em{\rm G}$}}
\def\IH{\relax{\rm I\kern-.18em H}}
\def\II{\relax{\rm I\kern-.18em I}}
\def\IK{\relax{\rm I\kern-.18em K}}
\def\IL{\relax{\rm I\kern-.18em L}}
\def\IM{\relax{\rm I\kern-.18em M}}
\def\IN{\relax{\rm I\kern-.18em N}}
\def\IO{\relax\hbox{$\inbar\kern-.3em{\rm O}$}}
\def\IP{\relax{\rm I\kern-.18em P}}
\def\IQ{\relax\hbox{$\inbar\kern-.3em{\rm Q}$}}
\def\IR{\relax{\rm I\kern-.18em R}}
\font\cmss=cmss10 \font\cmsss=cmss10 at 10truept
\def\IZ{\relax\ifmmode\mathchoice
{\hbox{\cmss Z\kern-.4em Z}}{\hbox{\cmss Z\kern-.4em Z}}
{\lower.9pt\hbox{\cmsss Z\kern-.36em Z}}
{\lower1.2pt\hbox{\cmsss Z\kern-.36em Z}}\else{\cmss Z\kern-.4em Z}\fi}
\def\IGa{\relax\hbox{${\rm I}\kern-.18em\Gamma$}}
\def\IPi{\relax\hbox{${\rm I}\kern-.18em\Pi$}}
\def\ITh{\relax\hbox{$\inbar\kern-.3em\Theta$}}
\def\IOm{\relax\hbox{$\inbar\kern-3.00pt\Omega$}}

\let\epsilon=\varepsilon
\ifx\answ\bigans \else\noblackbox\fi

\Title{\vbox{\hbox{UPR--549T}}}{\vbox{\centerline{Renormalization of
Chiral Couplings} 
\vskip2pt\centerline{in Tilted Bilayer Membranes}}}

\centerline{Philip Nelson and Thomas Powers}\smallskip
\centerline{Physics Department, University of Pennsylvania}
\centerline{Philadelphia, PA 19104 USA}
\bigskip\bigskip

We study the effects of chiral constituent molecules on the
macroscopic shapes attained by lipid bilayer membranes. Such fluid
membranes are beautiful examples of statistical ensembles of random
shapes, sometimes coupled to in-plane order. We analyze them with
methods of continuum elasticity theory, generalizing the well-known
Canham-Helfrich model, and in particular incorporate the effects of
thermal fluctuations.  The
condition that coordinate choice be immaterial greatly constrains the possible
forms of the statistical weights in these systems, leading to very few
independent couplings and hence physically simple models.
Thermal fluctuations effectively reduce the
chirality of a membrane at long scales, leading to an anomalous
scaling relation for the radius of bilayer tubules and helices as a
function of chirality. En route to this conclusion we develop a
perturbative calculation scheme, paying particular attention to the
functional measure needed to describe fluctuations covariantly.

\Date{4/93}\noblackbox                 


\def\ppro{\prod_\xi\prod_{a=1}^3}
\def\vx{\vec x\,}\def\vxb{\vec{\bar x}\,}\def\dvx{\dl\vec x}
\def\tr{\Tr}
\def\eff{_{\rm eff}}
\def\cost{{c_0^*}}
\def\gb{\bar g}
\def\dxi{\dd^2\xi}\def\gdxi{\sqrt g\dxi}\def\dgbxi{\sqrt{\bar g}\dxi}
\def\ddu{[\dd\dl u]}
\def\dk{{\dd^2k\over(2\pi)^2}}
\def\coub{\CO_\ub}
\def\ceff{{c^*\eff}}
\def\ubb{\ub_{12}}
\def\ko{{\kappa_0}}\def\mo{{\mu_0}}\def\mob{\bar\mu_0}\def\meff{\mu\eff}
\def\eps{\epsilon}


\lref\Sacka{E. Sackmann, H.-P. Duwe and W. Pfeiffer, Physical Scripta
{\bf T25} (1989) 107.\optional{[Great figures. Moduli for DMPC.]}}
\lref\Sackb{E. Sackmann, H.-P. Duwe, and H. Engelhardt, Faraday
Discuss. Chem. Soc., {\bf81} (1986) 281.\optional{[Measure $\kappa$ by
dynamical method to be 7\cdot 10^{-13} erg.]}}
\lref\Schnpriv{J. Schnur, private communication.}
\lref\Sel{J. V. Selinger
and J. M. Schnur, ``Theory of chiral lipid tubules,'' preprint (1993).}
\lref\BrSe{J. V. Selinger, Z.-G. Wang, R. F. Bruinsma, and C. M. Knobler,
``Chiral Symmetry Breaking in Langmuir Monolayers and Smectic Films,''
Phys. Rev. Lett. {\bf 70} (1993) 1139.}
\lref\deGennes{P. de\thinspace Gennes, Comptes Rendus (Paris) ser. II {\bf
304} (1986) 259.}
\lref\Schnsalt{ J. S. Chappell and P. Yager,
Biophys. J. {\bf 60} (1991) 952; M. Markowitz, J. Schnur, and A.
Singh, ``The 
influence of the polar headgroups of acidic diacetylenic phospholipids
on microstructure morphology and Langmuir film behavior;'' ``The
influence of the polar headgroups of acidic diacetylenic phospholipids
on their self-assembly behavior,'' preprints 1992.}%
\lref\LPR{W. Cai, T. Lubensky, J. Prost, and S. Ramaswamy, 
``Dynamics of Lyotropic Lamellar Phases, '' preprint 1992.}

\lref\Nakaa{N. Nakashima {\it et al.}, \cmp{``Helical superstructures
are formed from chiral ammonium bilayers,''} Chem. Lett. {\bf1984} (1984)
1709\optional{ [Twisted ribbons. Chain melting at 34C for helix phase,
33C for vesicular; complete disintegration even at just 35C. Sense of
helices depends on chirality of molecules. Direct precipitation
method. Very slow (days).]};
K. Yamada {\it et al.}, \cmp{``Formation of helical super
structure from single-walled bilayers by amphiphiles with
oglio-L-glutamic acid head group,''} Chem. Lett. {\bf1984} (1984) 1713.\optional{ [Wound
ribbons. Direct precipitation. Hypothesis of ribbons as precursors to tubules.]}}
\lref\Georger{J. Georger {\it et al.}, \cmp{``Helical and tubular microstructures
formed by polymerizable phosphatidylcholines,''}
J. Am. Ch. Soc.  {\bf109} (1987)
6169. \optional{[Discover direct precipitation method in
phosphatidylcholines (previous Yager-Schoen et al was
vesicle-conversion method). Formation is very rapid.
Tubules form at $T<38C$, which they identify with $T_m$.
Up to 1200 microns long. Multilamellar and unilamellar; so reject the
liposome mechanism and hypothesize helices as precursors. Helical rippling.
Diameter is monodisperse and const; length is polydisperse.]}}
\lref\NaAsKu{N. Nakashima, S. Asakuma and T. Kunitake, \cmp{``Optical
microscopic study of helical superstructures on chiral bilayer
membranes''} J. Am. Chem.
Soc. {\bf107} (1985) 509 \optional{[bilayers are in fluid state;
slow growth of helices; turn into tubules by widening of ribbon, not
change of pitch; stable only below $T_m$ of 34C; instant conversion
upon fast heating; sense depends on chirality]}\optional{[``the formation of
helices found to depend on chirality and chain-length''; also loss of
helicity at chain-melting; also precipitation above $T_m$ nothing;
also helices are precursors]}}%
\lref\Treanor{
R. Treanor and M. Pace, \cmp{``Microstructure, order, and fluidity of
a polymerizable lipid by ESR and NMR,''} Biochim. Biophys. Acta {\bf1046}
(1990) 1\optional{[This is DC$_{8,9}$PC, thermal method of tubule
preparation. Chirality is important; racemic doesn't make tubules.
``Despite the rigid external appearance of the tubes 
our findings indicate that the lipid bilayer retains fluid character
in the buffer solution on the time scale of the ESR experiment
($10^{-10}$ to $10^{-6}$ sec).'' But it's not so clear that `fluid'
here refers to translational order --- instead seems to mean that
certain segments of molecules are flopping around (far from headgroup)
while others are ordered (close to headgroup). Still that's some
evidence for in-plane fluidity since true crystalline order would
suppress all that flopping. ``Both NMR and ESR results are
consistent with regions of ordered and highly mobile components of the
tubule bilayers.'']}.}
\lref\Thomas{%
B. Thomas {\it et al.}, \cmp{``X-ray diffraction studies of tubules
formed from a diacetylenic phosphocholine lipid,''} in {\sl Complex
fluids,} Mat. Res. Soc. Symp.  Proc. {\bf248} (1992) 83.
\optional{[Shash says on the phone that there's no combination reflections, hence no interlayer
correlations, so UNLIKELY to have in-plane xtalline order because
these usually go together. Hence most likely is tilted-hexatic. Of
course this still doesn't mean fluctuations are important! ``Our
data indicate the tubules to be non-interacting sheets of bilayers in
the $L_{\beta'}$ phase [i.e. one of the tilted hexatics]... long-range
in-plane algebraic order plus chain ordering... phase transition at
36C.'' In the talk he said tilt inferred from thickness. And in-plane
correlations much greater than expected for isotropic fluid, but no
evidence for crystalline order.]}}%
\lref\SinBur{A. Singh {\it et al.}, \cmp{``Lateral phase separation based on
chirality in a polymerizable lipid and its influence on formation of
tubular microstructures,''} Chem. Phys. Lipids {\bf47} (1988) 135.}%

\lref\LuMacb{T. Lubensky and F. MacKintosh, ``Theory of `ripple'
phases of lipid bilayers,'' preprint 1993.}

\lref\Orlando{O. Alvarez, \cmp{``Theory of strings with boundary,''} Nucl.
Phys. {\bf B216} (1983) 125.}
\lref\Browicz{E. Browicz, Zbl. Med. Wiss. {\bf28} (1890) 625.}
\lref\HPP{G. Hinshaw, R. Petschek, R. Pelcovits, ``Modulated phases in thin
ferroelectric liquid-crystal films,''
Phys. Rev. Lett. {\bf60} (1988) 1864;
G. Hinshaw and  R. Petschek, ``Transitions and modulated phases in chiral
tilted smectic liquid crystals,'' 
Phys. Rev. {\bf A39} (1989) 5914.} 
\lref\LaSe{S. Langer and J. Sethna, ``Textures in a chiral smectic
liquid-crystal film,'' 
Phys. Rev. {\bf A34} (1986) 5035.} 
\lref\Meun{J. Meunier, ``Liquid interfaces,''
J. Phys. (Paris) {\bf 48} (1987) 1819.} 
\lref\SSRNO{R. Strey {\it et al.}, ``Dilute Lamellar and L$_3$ phases in the
binary water--C$_{12}$E$_5$ system,''
J. Chem. Soc. Faraday Trans. {\bf 86}(12) (1990) 2253.} 
\lref\HelPro{W. Helfrich and J. Prost, ``Intrinsic bending force in anisotropic
membranes made of chiral molecules,''
Phys. Rev. {\bf A38} (1988) 3065.} 
\lref\DiPiMe{S. Dierker, R. Pindak, and R. Meyer, Phys. Rev. Lett.
{\bf 56} (1986) 1819; W. Heckl {\it et al,} Eur. Biophys. J. {\bf14}
(1986) 11.\optional{[star defects with chiral molecules]}}

\lref\Kla{H. Kleinert, ``Thermal softening of curvature elasticity in
membranes,''
Phys. Lett. {\bf 114A} (1986) 263.} 
\lref\ForGab{D. Forster and A. Gabriunas, ``Capillary waves on an
$\eps$-dimensional interface,'' Phys. Rev. {\bf A23} (1980) 2627.}
\lref\WaZia{D. J. Wallace and R. K. P. Zia, ``Euclidean group as a
dynamical symmetry of surface fluctuations:  the planar interface and
critical behavior,'' Phys. Rev. Lett. {\bf 43} (1979) 808.}
\lref\LTR{T. C. Lubensky, T. Tokihiro, and S. R. Renn, ``Polymers
dissolved in a chiral liquid crystal:  model for twist-grain-boundary
phases,'' Phys. Rev. {\bf A43} (1991) 5449.}
\lref\Helfb{W. Helfrich, ``Helical bilayer structures due to spontaneous
torsion of the edges,''
J. Chem. Phys. {\bf 85}(2) (1986) 1085.} 
\lref\DaLe{F. David and S. Leibler, ``Vanishing tension of fluctuating
membranes,'' J. Phys. II France {\bf 1} (1991) 959.} 
\lref\PePr{L. Peliti and J. Prost, ``Fluctuations in membranes with reduced
symmetry,'' J. Phys. France {\bf 50} (1989) 1557.} 
\lref\SMMS{See for example {\sl Statistical mechanics of membranes and
surfaces,} D. Nelson {\it et al.}, eds (World Scientific, 1989).}
\lref\DSMMS{F. David, in \SMMS.}
\lref\OYL{Ou-Yang Zhong-Can and Liu Jixing, ``Theory of helical structures
of tilted chiral lipid bilayers,''
Phys. Rev. {\bf A43} (1991) 6826.} 
\lref\Can{P. Canham , J. Theor. Biol. {\bf26} (1970) 61}
\lref\Helfaa{W.
Helfrich, Naturforsch. {\bf28C} (1973) 693\optional{[introduces tilt
by analogy with lyotropic l.c.'s!]}.}
\lref\NePe{D. Nelson and L. Peliti, ``Fluctuations in membranes with
crystalline and hexatic order,'' J. Phys. {\bf 48} (1987) 1085.}
\lref\DaLa{I. Dahl and S. Lagerwall,
Ferroelectrics {\bf 58} (1984) 215.}
\lref\DeTa{P.G. de\thinspace Gennes and C.  Taupin, J. Phys. Chem.
{\bf86} (1982) 2294.}
\lref\DaGuPe{F. David, E. Guitter, and L. Peliti, ``Critical
properties of fluid membranes with hexatic order,'' J. Phys. (Paris)
{\bf48} (1987) 1085.}
\lref\NePelc{D.R. Nelson and R. Pelcovits, 
``Momentum-shell recursion relations, anisotropic spins, and liquid
crystals in $2+\epsilon$ dimensions,''
Phys. Rev. {\bf B16} (1977) 2129.}

\lref\Davflow{F. David, Europh. Lett. {\bf6} (1988) 603.}  
\lref\tether{Y. Kantor, M. Kardar, and D.R. Nelson, Phys. Rev. Lett. {\bf
57}, 791 (1986); Phys. Rev.  {\bf A35} 3056 (1987); {\it ibid.}  {\bf A38}
4020 (1987); M. Paczsuki, M. Kardar, and D.R. Nelson, Phys. Rev. Lett. {\bf
60}, 2638 (1988).}
\lref\Ratpriv{B. Ratna, private communication.}
\lref\JPAS{J. Polchinski and A. Strominger, ``Effective string theory,''
Phys. Rev. Lett. {\bf67} (1991) 1681.}
\lref\Polbook{A. Polyakov, {\sl Gauge fields and strings}, (Harwood,
1987).}

\lref\PNTRP{P. Nelson and T. Powers, Phys. Rev. Lett. {\bf69} (1992) 3409.}

\lref\LeiAnd{S. Leibler, J. Phys. {\bf47} (1986) 507;
S. Leibler and D. Andelman, J. Phys. (Paris) {\bf48}
(1987) 2013.\optional{[curvature instability gives rippling]}}%
\lref\GolLei{R. Goldstein and S. Leibler, Phys. Rev. Lett.
{\bf61} (1988) 2213.}%
\lref\LeiSMMS{S. Leibler, in \SMMS.}
\lref\Ratna{A. Rudolph, B. Ratna, and B. Kahn,
\optional{[``Self-assembling phospholipid filaments'']} Nature {\bf352}
(1991) 52.}%
\lref\LuProst{T.C. Lubensky and J. Prost, ``Orientational Order and
Vesicle Shape," submitted to J. Physique (Paris).}
\lref\NePel{D. Nelson and L. Peliti, J. Phys. {\bf 48} (1987) 1085.}
\lref\bouligand{F. Livolant and Y. Bouligand, J. Physique (Paris) {\bf 47},
1813 (1986).}%
\lref\DaGuPe{F. David, E. Guitter, and L. Peliti, J. Phys. {\bf 48}
(1987) 2059.\optional{[hexatic membranes may undergo crumpling transition]}}%
\lref\Pomeau{Y. Pomeau and J. Lega, ``Structures macroscopiques en
spirales comme configuration d'\'equilibre d'ensemble de mol\'ecules
chirales,'' C.R. Acad. Sci. Paris {\bf II:311} (1990) 1135; N.
Mendelson, Comm. Theor. Biol. {\bf1} (1989) 217. 
\optional{[spiral structures in mutant bacteria]}
}
\let\french=\Pomeau
\lref\GoLu{L Golubovi\'c and T. C. Lubensky, \cmp{``Smectic elastic
constants of lamellar fluid membrane phases: crumpling effects,''}
Phys. Rev. {\bf B39} (1989) 12110.}
\lref\techno{A. Rudolph, J. Calvert, P. Schoen, and J. Schnur,
``Technological development of lipid based tubule microstructures,''
in {\sl Biotechnological applications of lipid microstructures,} ed.
B. Gaber {\it et al.} (Plenum, 1988).}
\def\chistar{\DiPiMe\ref\achistar{X. Qiu {\it et al.}, Phys. Rev. Lett.
{\bf67} (1991) 703. \optional{[star defects with achiral molecules]}}}
\def\helixref{\Nakaa\Georger}
\def\tubuleref{\ref\YaSh{P. Yager and P. Schoen,
\cmp{``Formation of tubules by a polymerizable surfactant,''} Mol. Cryst.
Liq. Cryst. {\bf106} (1984) 371.\optional{[This is the old
liposome-conversion method, imperfect, multilamellar.]}}}
\let\stria=\Sel
\lref\bigkap{C. Safinya, ``Rigid and fluctuating surfaces,'' in {\sl
Phase transitions in soft condensed matter,} ed. T. Riste and D.
Sherrington (Plenum, 1989).\optional{[usually kappa is 20-40 kT, but
here it's approximately equal to kT.]}}
\def\steric{\ref\Helfster{W.  Helfrich, Z. Naturforsch. {\bf 33a}
(1978) 305.}\bigkap}
\def\runkap{\DeTa--%
\nref\Helfa{W. Helfrich, J. Phys. (Paris) {\bf46} (1985) 1263; {\it
ibid.} {\bf47} (1986) 321.}%
\nref\PeLe{L. Peliti and S. Leibler, ``Effects of thermal fluctuations on
systems with small surface tension,''
Phys. Rev. Lett. {\bf54} (1985) 1690.}
\nref\Fors{D. F\"orster, ``On the scale dependence, due to thermal
fluctuations, of the elastic properties of membranes,''
Phys. Lett {\bf114A} (1986) 115.}%
\nref\Pold{A. Polyakov, ``Fine structure of strings,''
Nucl. Phys. {\bf B268} (1986) 406.}
\Kla}
\def\noshear{\Sacka}
\lref\Kavalov{A. Kavalov,  ``Model of random surfaces with long distance
order,'' preprint 1992.}
\lref\annoying{A. Singh, P. Schoen, J. Schnur, Chem. Commun. {\bf
1988} (1988) 1222
\optional{[Achiral molecule makes old-style tubules
(vesicle conversion, wrapping). Helices don't form. ]}}
\lref\Mavra{N.E. Mavromatos, J.L. Miramontes, ``Regularizing the
functional integral in 2-D quantum gravity,''  Mod. Phys. Lett. {\bf
A4} (1989) 1847; E. D'Hoker, ``Equivalence of Liouville theory and 2-D
quantum gravity,'' Mod. Phys. Lett. {\bf A6} (1991) 745.}
\lref\DDK{F. David, ``Conformal field theories coupled to 2-D gravity in the conformal
gauge,''
 Mod. Phys. Lett. {\bf A3} (1988) 1651; J. Distler and H. Kawai,
``Conformal field theory and 2-D quantum gravity,''  Nucl. Phys. {\bf
B321} (1989) 509.}
\lref\MFM{G. Moore and P. Nelson, \cmp{``Measure for moduli,''} Nucl. Phys.
{\bf B266} (1986) 58.}
\lref\Polch{J. Polchinski, \cmp{``Evaluation of the one loop string path
integral,''} Commun. Math. Phys. {\bf 104} (1986) 37.}
\def\tiltex{\ref\Rhodes{D. Rhodes {\it et al.}, Chem. Phys. Lipids
{\bf49} (1988) 39.\optional{[Tilt 28$^\circ$. DC8,9PC. Direct
precipitation method.]}
}}


\newsec{Introduction and Summary}

Amphiphilic molecules in water can self-assemble into large, stable sheets
a few tens of \AA ngstroms thick. These sheets in turn can arrange themselves
into a wide variety of beautiful stable structures, often with enormous 
characteristic
scales (on the order of microns or larger) \SMMS. The problem of
understanding the structures which arise sits at a juncture between physics
and biology: on one hand bilayer membranes are simplified analogs of
important biological membranes, and their transformations are reminiscent of
biological processes (\eg\ vesiculation, echinocytosis, \etc). On the other
hand, it has become clear that some of the behavior of these simple analog
systems can be understood on the basis of simple physical models and
statistical mechanics. As we recall below, much of the effective simplicity
comes from the large discrepancy of scales between the constituents and the
structures they form, a familiar phenomenon in physics. So it seems that
physics and biology each stand to gain some ideas from the other.\foot{The
study of bilayer membranes also has a host of potential technological
applications; for references see \techno\OYL.}

The physics of bilayer membranes has been studied extensively \SMMS. In this
paper, we will study the effects on membranes of chirality, thermal
fluctuations, and especially the interplay of these. Our results were
described briefly in \PNTRP.

``Chirality'' refers to the influence of constituent molecules which are not
the same as their mirror images. A theory describing structures built from
such molecules must explicitly break parity symmetry in a way we will
recall below. Such symmetry breaking should, in principle, be very common
in nature, since many of the lipid molecules commonly studied are, in fact,
chiral. Still, one may wonder whether this qualitative fact can affect
the very large (micron scale) structures formed from membranes. After all,
whenever we have a large discrepancy of length scales in physics, the
long-scale behavior forgets most of the features of the constituents,
remembering only a few phenomenological parameters. Why should chirality
be one of the properties remembered?  Moreover, in the
molecules in question chirality is a very minor aspect of molecule shape.
Why should it matter?

In fact, sometimes chirality does {\it not} matter for the shapes attained
by lipid bilayers, while other times it is crucial.\foot{Several interesting
chiral effects also occur in monolayers, notably the formation of striped
phases \LaSe\HPP,  star defects \chistar, and asymmetric rippled
phases~\LuMacb.} We recall in Section~2 some
of the relevant experimental facts. Here we simply note that in many
instances chiral behavior stops abruptly as one raises the temperature
through a critical value comparable to the chain-melting temperature $T_m$.
We will argue in Section~3 that this is no accident: the presence of {\it
tilt} order is crucial for chirality to express itself on long scales.
The argument hinges on the phenomenon of ``accidental symmetry'' well
known in particle physics: above $T_m$ there are simply no relevant or
marginal effective interactions consistent with the required symmetries which
could communicate chirality to long scales. With tilt order, however,
chirality becomes strongly relevant with spectacular effects such as the
formation of large helical ribbons \helixref. Remarkably there is just {\it
one} allowed bulk chiral term, plus a total derivative term, leading to
rather simple physics.

The above considerations are implicit in previous work \HelPro\PePr\OYL,
but we will give what we believe to be a very simple route to the conclusion,
with the added benefit that it becomes easy to enumerate all allowed couplings,
including some missed by the other methods. Our strategy is to find all the
required symmetries of our system, including a subtle discrete symmetry, by
describing the configurations in a purely geometrical form. We will also
give a simple account of the rule for counting dimensions, and discuss
why it differs from the one appropriate to the Polyakov string theory,
leading to interactions not anticipated in that theory.\foot{In
particular our chiral term is not the Wess--Zumino term introduced in
\Kavalov.} 

Thermal fluctuations have long been known to be important in bilayer
membranes. For example, they cause gigantic undulations in the shape
of red blood cells (the ``flicker'' phenomenon \Browicz\SMMS). As we will recall,
thermal fluctuations are determined by a dimensionless phenomenological
parameter, $k_BT/\kappa_0$, whose value is generally around $1/40$,
but can be as large as unity for suitably prepared systems \bigkap.
Fluctuation effects are also enhanced when there is a large range of scale,
as in our problems, due to large logarithms. In any case, the stiffness,
$\kappa_0$, while not infinite, is generally large enough to justify a
perturbative expansion about the fixed point with $k_BT/\kappa_0=0$. We
also expand in powers of the chirality $\cost$, since as we will see it is
quite small compared to the cutoff. Fluctuations in the in-plane order can
also be important, certainly near the temperature $T_m$ where this order
is destroyed. We will work below this temperature, however, so that all
fluctuations are weakly coupled.

Thermal fluctuations give rise to several striking phenomena, for example
the well-verified steric effective interaction \steric. Equally striking,
but harder to verify experimentally, is the prediction of a scale-dependence
for the effective stiffness $\kappa\eff (L)$ on scale $L$ \runkap. In this
paper we will find a related scale dependence in the effective value of the
chiral coupling $\ceff$, and argue that this dependence may be easier
to observe than that of $\kappa\eff$.

To calculate the effects of fluctuations we need a functional measure
on the space of configurations, consistent with the symmetries of
Section~3. In Section~4 and Appendix A we review this problem: while
extremely difficult in general, it becomes quite easy to lowest
nontrivial order in our expansion about infinite stiffness $\kappa_0$.
We discuss carefully the gauge-fixing correction required for
``Monge'' gauge, and show that contrary to other claims it is not
trivial. Fortunately, this factor does not affect the renormalizations
of $\kappa_0$ and $\cost$ to one loop, but as we point out it is necessary in order
to reconcile the renormalization of the area coefficient $\mu_0$ in Monge gauge
with the normal-gauge result.

In Section~5 we finish setting up our perturbation theory and discuss
carefully the issue of field renormalization. Finally, in Section~6
we compute the scale dependence of our bulk chiral coupling $\cost$ to one
loop and find that like $\kappa_0$, it suffers a logarithmic renormalization.
While this effect may seem no easier to verify than the elusive running of
$\kappa_0$, we point out in Section~7 that it should have a simple
consequence: the radii $R$ of helices and  tubules
should obey an anomalous scaling law as we dilute the chiral amphiphiles
with similar achiral molecules. Unfortunately, we cannot predict the
exponent of this law which is nonuniversal. We do give the exponent in terms
of the effective stiffness, $\kappa\eff$, which may be independently
measurable. The relation we get is
$$
R\propto (\epsilon)^{-(1+k_BT/4\pi\kappa\eff)}
$$
where $0<\epsilon<1$ is the dilution fraction and $\kappa\eff$ is the
effective stiffness.

\newsec{Some Experimental Facts}

Often chirality does not appear to matter in bilayer membranes, which
frequently form lamellar or vesicular phases with no obvious gross
chiral character.
A noteworthy exception is the phase in which helices and/or tubules form.
The former appear to be ribbons of bilayer of constant width and very long
length, ``wrapped'' around an imaginary cylinder with a definite handedness
\helixref. Tubules appear as long, uniform cylinders, either unilamellar
or up to 20 layers thick, with radius up to a micron and almost arbitrarily
long\foot{At least two kinds of tubules are known, with
apparently different formation mechanisms. Those described in
\tubuleref\ seem to form from wrapping of liposomes; tubules of this
sort do not require chiral constituent molecules \annoying,
and we have nothing to say about them. The ones we will discuss \helixref\
precipitate directly from solution; like the helices, they only seem
to form from chiral molecules.}%
. A very broad class of amphiphiles have been found which
readily form helices and tubules.

While tubules may not at first seem chiral, they usually display under
the electron microscope helical rippling \Georger\ or (when plated
with metal)  a single helical striation of definite
handedness \stria. Also, in some cases the helices are clearly precursors
to the unilamellar tubules \Nakaa\Georger. We will consider both helices and tubules
as paradigms of chiral structures.\foot{It is possible that other
twisted-ribbon structures of greater biological relevance can also be
understood by techniques such as ours \french.} In each case a key observation
is that when chiral structures form, they all have the same handedness,
which reverses when one starts with the opposite constituent molecules
\Nakaa\NaAsKu. Thus, the system certainly remembers the chirality of its
constituents on some occasions (tubules, helices) but not on others
(vesicles, lamell\ae). What distinguishes these cases?

A key observation is the role played by temperature. Helices and tubules
typically cannot form above a temperature variously reported to be
between $34^\circ$C and $38^\circ$C; once formed,
tubules also disintegrate 
suddenly as the temperature is raised through this special value
\Nakaa\Georger\NaAsKu.
As mentioned in the Introduction, for each material
this special temperature is close to
the chain-melting temperature $T_m$. Below $T_m$ we may think of the molecules
as having long straight tails, locally pointing in the same direction
for optimal packing (\tfig\fconfig a). Typically this direction is tilted
away from the normal to the plane, again for packing reasons. The polar angle
of tilt is fixed, typically near $30^\circ$ \tiltex, 
while the azimuthal angle is an 
angular order parameter represented by $\hat m$ in \fconfig c. 
Above $T_m$ the chains are disordered.

\ifigure\fconfig{Successive idealizations of the configuration space: (a)
cartoon; (b) Nematic vector field at fixed angle to the normal; (c)
in-plane unit tangent vector field $\hat m$ obtained by choosing a
normal $\hat n$ and dropping a perpendicular from the director to the
tangent plane.}{fconfig.eps}{3.1}

Evidence for tilt order in tubules comes from several experimental
measurements and theoretical prejudices. First, x-ray scattering on
(hydrated) tubules shows a layer thickness compatible with a nonzero
tilt~\Thomas. Furthermore, monolayer studies of many diacetylenic 
phospholipids show that some do not have an in-plane ordering transition near
$35^\circ$C; none of these lipids form tubules, while most of the ones
which {\it do} exhibit a transition {\it do} form tubules~\Schnpriv. 
Also, a number of measurements indicate some degree of
in-plane order, but {\it crystalline} order seems ruled out \Treanor\Thomas\Schnpriv.
These observations fit with our expectations for an infinite, isolated
membrane, where dislocations eliminate crystalline order but orientational
quasi-long-range order survives up to a Kosterlitz-Thouless type transition
\NePe. The lack of crystalline (translational) order also follows
experimentally from the
shear modulus, which is essentially zero \noshear.\foot{Of course there is a
question of time scales here. What we are really asserting is that the
rate for molecules to diffuse around each other, relieving shear stress, is
fast compared to the characteristic frequencies of the modes of interest
to us.}

Finally, the interactions responsible for bilayer membrane shapes
appear to be {\it local}. In contrast, de\thinspace Gennes has
proposed a theory of tubule formation based on a nonlocal,
electric-moment force between bilayer edges \deGennes. This model
predicts a large effect of solvent salinity on tubule formation, as
the electrostatic force gets screened; experiments do not see such a
clear effect \Schnsalt. We will work under the assumption that all
forces are local, \ie, any nonlocal interactions are screened to a
scale smaller than the one we study. We will also neglect the effects
of self-avoidance; while local in three-space, such effects look
nonlocal from the two-dimensional point of view. Since our membranes
are fairly rigid, they will not self-intersect and we can ignore this
complication.

To summarize: lipid bilayers seem to have a low-temperature phase with
orientational quasi-long-range order carried by tilt (we may assume that
hexatic order, if present, is locked to the tilt order) but no translational
order. Hence,  these layers resemble smectic-$C^*$ liquid crystals. Their
stiffness, $\kappa_0/k_BT$, while large, is not extremely large \Sacka.
Finally, the chirality of bilayers will be seen in the next section to be
described by an interaction with a phenomenological coupling constant
$\cost/k_BT$ with the dimensions of inverse length. Since, as we will
recall in Section~7, the radius of helices (and presumably tubules as
well) $R$ is proportional to $\ko/\cost$ in
mean-field theory \HelPro, and $R$ is known to be enormous compared to the
cutoff $\Lambda^{-1}$, it follows that experimentally $\cost$ is small.
This accords with our intuition that chirality is a very minor feature in
the shape of these amphiphiles. We will accordingly set up in Sections~4--6
a perturbation expansion in powers of $\cost$ and $\ko^{-1}$.

The above remarks could also apply to the asymmetric rippled lamellar
phases studied in \LuMacb. The techniques in this paper could be taken
over to study the effects of fluctuations in these systems as well as
tubules and helices.

\newsec{Modes and Couplings}

When trying to understand physics at scales much larger than the constituent
size, we can neglect the discrete character of the constituents and write
a continuum model. Moreover, of the many degrees of freedom in the
original problem, only a few will be important for the long-scale physics
and only a few couplings between these modes are needed. The remaining
modes can be integrated out of the path integral and their effects
summarized by renormalizations of the few retained couplings. Let us
recall the general principles and apply them to our system.

\subsec{Modes}

The important modes for equilibrium statistical mechanics will be those
arising from the spontaneous breakdown of a continuous symmetry (elastic
modes), plus any modes having all-scale fluctuations due to a phase transition.
An example of the latter would be the average polar angle of tilt close
to the temperature where tilt order is lost. Since we will only consider
systems far from phase transitions, we will take our model system to have
only elastic modes.

Superficially the counting of elastic modes looks straightforward. A
featureless flat 2d surface breaks one translational symmetry in 3-space,
leading to one soft mode, the ``undulation'' mode. Two-dimensional crystalline
order, if it were present, would break the other two translation symmetries
and lead to the two in-plane ``sound'' modes familiar from tethered membrane
theory. In our case there is no such order, but {\it orientational} in-plane
order breaks one symmetry, that of rotations in the plane.\foot{The other
two rotations out of the plane are already correctly accounted for by the
undulation mode.} Thus, when tilt order develops we get a second elastic
mode, the azimuthal angle of the average tilt. The polar angle of tilt
is not associated to any symmetry, and we can take it to be constant
at fixed temperature.\foot{Near a defect 
we would have to consider fluctuations of the polar angle as well.}
Hexatic order can form in addition to tilt, but in the absence of tilt
we will see that it cannot lead to the expression of chirality, while
in the presence of tilt it does not lead to a new elastic mode, but
simply locks to the tilt order parameter. Thus we will not include an
explicit hexatic order parameter.

While we will accept the conclusion just reached, we need to add some
remarks before it becomes convincing. First, we have so far acted as
though our statistical sum were over {\it configurations} of the membrane,
when of course we should instead sum over {\it phase} space. We can
imagine integrating out the momentum modes, leaving a sum over configurations
only, but in general this could lead to new long-range interactions.
Fortunately David has shown that thermal undulations have the effect of
screening such interactions down to short range, so their effects may
be incorporated into renormalizations of the local interactions we will
consider \Davflow.

Second, the description of a fluid membrane as a mathematical
two-surface leaves no room to discuss compression or rarefaction of
the molecules in the plane. We could rectify this by introducing a
local density variable, but the energy of its fluctuations (controlled
by the compression modulus) is very large on long length scales
compared to the modes we will keep; we do not get a soft mode by
passing to long wavelengths since no symmetry is being
broken.\foot{For {\it dynamics} we certainly do need this mode
since it is associated to a conserved quantity~\LPR.}

Physically, to a given mathematical 2-surface we can associate many
configurations by ``tiling'' the surface with molecules of fixed area.
Since all we really measure is the overall shape, we will let the abstract
surface represent all such configurations, writing down the most general
local free energy functional consistent with the microscopic system's
symmetries to take account of all the degrees of freedom discarded. The
only way this can fail is when the omitted modes are elastic and hence
capable of generating long-range interactions; we account for this by
explicitly including a tilt order parameter.

Thus we are led to a continuum description of our important long-scale
degrees of freedom: a configuration of our system is a two-surface in
3-space; on this surface is given a field of headless (nematic) vectors
of unit length, making a fixed angle with the normal (\fconfig b). An equivalent
and more convenient description is shown in \fconfig c. Since only the azimuthal
angle of tilt matters, we can represent the director at $\xi$ by projecting
it down to the tangent plane and scaling to unit length. This gives a unit
vector field $\hat m (\xi)$ describing tilt. It is very
important to note that this order parameter is not an abstract phase,
as in superfluids, but rather a {\it tangent} vector to a curved surface.

A word of caution is
needed concerning $\hat m$: since the bilayers we consider are
symmetric, they have no 
preferred choice of normal, or equivalently, no preferred
orientation.\foot{A choice of normal is the same thing as a choice of
orientation because we {\it do} have a preferred orientation for
3-space. The lack of preferred orientation prevents us from adding
a Wess-Zumino term of the sort proposed by Kavalov \Kavalov.}
To obtain $\hat m (\xi)$ given the data in \fconfig b we must first {\it
choose} a normal $\hat n$; the other choice $-\hat n$ will describe the
same configuration with the vector field $-\hat m$. Nevertheless the
description in \fconfig c will prove convenient, as we will need to choose
$\hat n$ anyway for other reasons later.

Hence, our final description of configuration space is the following: a
configuration is a 2-surface in 3-space with a choice of normal vector field
$\hat n$ and a unit tangent vector field $\hat m$. Two surfaces
related by $\hat n' =-\hat n$, $\hat m' =-\hat m$ are to be regarded as
identical.

\subsec{Symmetries}

Having found the important elastic degrees of freedom, we now need to specify
their interactions. The general rule is to write the most general
local Hamiltonian
functional $H$, retaining only terms of low dimension (\ie\ only relevant
and marginal terms) respecting the appropriate symmetries. Giving each such
term its own unknown coefficient we can then calculate many things by doing
path integrals with statistical weight $[\dd \bar\mu]\ex{-\beta H}$, where
$[\dd \bar\mu]$ is some appropriate local measure. We will discuss this
measure in Section~4 and Appendix~A; here we will consider the free energy
functional $H$. We will take a purely two-dimensional point of view
throughout; that is, we will not attempt to derive our couplings from those
of 3d liquid crystals. This will make it easy to enumerate all possible
terms; we will find some terms missed by previous authors.

Before we can write $H$ we need a concrete way to describe the geometrical
data of a configuration in terms of ordinary functions. To get this we need
to make some further choices analogous to the choice of $\hat n$ above: we
need to specify local coordinates $\xi^i$, $i=1,2$. Neglecting tilt
for the moment,
our surface can then be
described in many ways by giving three functions $x^a(\xi^i)$, $a=1,2,3$.
Any reparameterization changes these functions without affecting the physical
configuration, so when we write $H=H[x^a(\xi^i)]$ we must demand that $H$
be reparameterization invariant. Physically this reflects the fact that
since the molecules can slip past each other in the plane, we cannot
permanently affix any pair of coordinates to each individual molecule.
(In the case of tethered membranes each constituent {\it is} assumed to
be permanently linked to its neighbors and we need not enforce
coordinate invariance.)

Once we have coordinates we can express the vector field $\hat m$ by its
components $m ^ i$. We can also readily express the restriction of locality:
$H$ must be an integral over $\dd^2\xi$ of some function depending on
$\vec x$ and $\hat m$ only via the values $x^a(\xi)$ and $m^i(\xi)$ and
their derivatives, evaluated at $\xi$. Since a small change of $\xi^i$
moves us a short distance in 3-space, this condition indeed expresses
that only near neighbor molecules interact.\foot{We will not consider
self-avoidance in this paper.}

The two components $m^i$ are not independent, since $\|\hat m\|^2=1$. Later
we will find it convenient to write $\hat m$ in terms of a single
angular function $\theta(\xi)$ as follows:
\eqn\emth{\hat m = \hat e_1 \cos\theta + \hat e_2 \sin\theta \;.}
Here $\hat e_\alpha(\xi)$, $\alpha=1,2$ are a pair of orthonormal tangent
vector fields. The $\hat e_\alpha$ are not additional dynamical variables
but rather an additional unphysical choice analogous to the choice of $\hat n$
and $\xi^i$. That is, we first must sum over surfaces. For every surface
we {\it choose} a frame field $\{\hat e_1,\hat e_2\}$. Then we sum over all
$\theta(\xi)$. Thus we have three redundancies in our description, leading
to three symmetry requirements on $H[x^a,\theta]$: the discrete symmetry
$\hat n\mapsto -\hat n$, $\hat m\mapsto -\hat m$, 2d coordinate invariance,
$ \xi^i\mapsto\xi^{i\prime}=f^i (\xi)$, and change of orthonormal frame,
$\hat e_\alpha\mapsto\hat e_\alpha '=R_\alpha{}^\beta(\xi)\hat e_\beta(\xi)$
where $R$ is some rotation matrix depending on position. Finally, in addition
to these redundancies we have the physical (active) symmetries of rigid
Euclidean motions, $E_+(3)$, which actually change the original geometrical
configuration. The plus sign reminds us that we do {\it not} enforce
symmetry under 3-space parity transformations.

To find the most general expression for $H$ consistent with these symmetries,
we construct from $\vec x$ and $\hat m$ a few building blocks with simple
transformation laws under 2d coordinate change. The $x^a$ themselves transform
as three scalars. Any vector field $\vec A$ on our surface will also have
components $A^a(\xi)$, $a=1,2,3$ transforming as scalars. However, if
$\vec A (\xi)$ is known to be {\it tangent} to the surface this is a
redundant description: two components should suffice. We can introduce
a basis of tangent vectors $\vec t_i=\partial_i \vec x$, $i=1,2$ and
expand $\vec A=A^i \vec t_i$; the $A^i$ now transform as 2-vectors under
coordinate change. We can also construct the second-rank tensor $g_{ij}=
\vec t _ i \cdot \vec t _j$ and its determinant $g\equiv\det[g_{ij}]$.
{}From the metric we then get an invariant volume element $\sqrt{g}\dd^2\xi$
as well as a covariant derivative rule $\nabla_i$. We will recall the
definition of $\nabla$ when we choose a gauge in Section~4, but for now note
one key property: two covariant derivatives $\nabla_i$, $\nabla_j$ {\it
commute} when acting on a scalar. Thus the second-rank tensor $K_{ij}=
\hat n\cdot (\nabla_i\nabla_j \vec x)$ is symmetric. This tensor clearly
vanishes when $\vec x$ is a linear function of $\xi$ (flat surface); $K_{ij}
(\xi)$ measures the curvature of our surface, its tendency to bend away from
its tangent plane at $\xi$ as we move away from $\xi$.

We need one further geometrical construction. Let $\eps^{abc}$ denote the
antisymmetric tensor with $\eps^{123}=+1$. From this we can construct a third
tensor on our surface:
\eqn\ealter{\eps_{ij}\equiv t_i{}^a t_j{}^b \hat n^c \eps^{abc}\;.}
Unlike $g_{ij}$ and $K_{ij}$, $\eps_{ij}$ is antisymmetric.\foot{To
relate this to the usual definition we note that if $\xi^i$ have the
orientation induced by our choice of $\hat n$ then \ealter\ says that
$\epsilon_{ij} = \sqrt g{0\ \ 1\choose{\rm -}1\  0}_{ij}$.}

Given a surface and a choice of normal $\hat n$, we have thus constructed
a density $\sqrt{g}\dd^2\xi$, a covariant derivative $\nabla_i$, and three
tensors $g_{ij}$,$K_{ij}$, and $\eps_{ij}$ on the surface. Each of these
ingredients has no dependence on our choice of coordinates other than the
one implied by the 2-space indices, so it is easy to construct
coordinate-invariant expressions from them. Each also behaves simply under
change of normal: under $\hat n\mapsto -\hat n$, $K_{ij}$ and $\eps_{ij}$
change sign. Finally, each has been constructed to be invariant under all
proper Euclidean motions (rigid translation and rotation in 3-space). Thus
any $H$ constructed from them and the tilt vector field $\hat m$ will have
all the required symmetries provided that all
$i,j\ldots$ indices are contracted and we enforce symmetry under
\eqn\ztwo{
\hat m\mapsto -\hat m\;,\qquad
K_{ij}\mapsto -K_{ij}\;,\qquad
\eps_{ij}\mapsto -\eps_{ij}\;.}

\subsec{Dimensions}

We now need to associate naive scaling dimensions to these ingredients. Near
the weakly-fluctuating limit $\kappa\to\infty$ these will be close to the
true scaling dimensions. Our key observation is that the scale transformations
of interest involve {\it physical} 3-space. Scale changes in the {\it internal}
coordinates $\xi^i$ are just as unphysical as any other coordinate
transformation; every term of $H$ will be trivially invariant under such
transformations. To count powers we thus note that $\vec x$ has length dimension 1 while the unit
vectors $\hat n$, $\hat m$ are dimensionless. Derivatives with respect to
$\xi^i$ have no 3-space dimension, so $g_{ij}$ and $\sqrt g$ have dimension
2 while $K_{ij}$ has dimension 1. The inverse metric $g^{ij}$ has dimension
$-2$, so $\tr K\equiv g^{ij}K_{ij}$ has dimension $-1$. The Laplacian $\Delta
=g^{ij}\nabla_i\nabla_j$ has dimension $-2$. Finally, the alternating tensor
$\eps_{ij}$ \ealter, has length dimension 2.

Putting things together, we find for example that $\int\sqrt{g}\dd^2\xi
(\tr K)^2$ has length dimension zero, so its coupling constant $\ko\over2$
is an energy and $k_BT/\ko$ is a pure number as claimed earlier.

These rules are so important, and so counterintuitive for
string theorists, that we should pause to elaborate them. We first note that
the physical probes we use to study membranes, and the imposed extensive
quantities, are characterized by length scales in 3-space (wavelength of
scattered light, total surface area, lamellar spacing, \etc). Also the
physical short-range cutoff $\Lambda^{-1}$ is the size of our molecules,
again a 3-space distance.\foot{Of course these facts have their analogs in
string theory; the invariance under coordinate transformation at fixed
3-space cutoff is incorporated into the Virasoro Ward identity, which in
turn gives rise to the Liouville measure factor \Polbook. We will mention
briefly later on how this factor enters our scheme. But the original, naive
dimension counting used to determine which terms to retain in $H$ proceeds
by the rules of ordinary 2d field theory, \ie\ using an ordinary cutoff
on $\xi$.  The difference comes from the different
fixed points controlling the physics in each case.} 

Secondly, our rules correctly predict the short-distance divergence
structure of Feynman diagrams as follows. Consider a scalar field
$\varphi$ with Hamiltonian of the schematic form $H=\int\dxi\bigl[(
\Delta\varphi)^2 + \sum_i\la_i\pa^{k_i}\varphi^{v_i}\bigr]$, where
$\la_i$ are small couplings. A Feynman graph with $L$ loops, $E$
external legs, and $n_i$ vertices of type $i$ will then have a
superficial degree of divergence $\dl=2+E +\sum_i n_i\dl_i$ where
$\dl_i=k_i-v_i-2$. In our case the role of $\varphi$ is played by
$\vec x$, and we indeed see that $\dl_i$ is just the length dimension
of the coupling $\la_i$ once we include factors of $g^{ij}$ and $\sqrt
g$ needed for covariance. We will see this divergence structure at
work in our explicit calculations in Section 6. 

Finally, the general
principle of naturalness says that in a local theory the physical values
of bare couplings should be on the order of an overall energy scale times
the appropriate powers of the cutoff.\foot{The couplings $\cost$ and
$\mu_0$ below the will prove to be smaller than this for reasons we will
describe.} For membranes the scale is the condensation energy per molecule,
about one electron volt, and indeed $\ko$ is comparable
to this~\Helfaa\Sacka. The corresponding stiffness to Gaussian
curvature, $\bar\kappa_0$, is hard to measure, but a simple mechanical
model of the membrane shows that it is comparable to $\kappa_0$
\LeiSMMS, as predicted by our dimension counting. In the counting 
appropriate to the Polyakov string, on the other hand, $\bar\kappa_0$
and $\ko$ have different dimensions.


\subsec{Enumeration of couplings}

We can now enumerate all allowed local couplings, naively relevant or marginal,
among our long-scale effective degrees of freedom. Our 2d covariant
approach makes it easy to be systematic.
Euclidean invariance requires that the
embedding $x$ enter the free energy functional only through the
Euclidean invariant quantities $K_{ij}$ and $g_{ij}$ or through an 
integrand that changes under a translation by a quantity that
integrates to zero.  An example of the latter is the volume term 
$\int\dd^2\xi\sqrt{g} x\cdot\hat n$, which is Euclidean invariant in
the case of a closed vesicle.  This term lacks the 
$\hat n \mapsto -\hat n$ symmetry required for a symmetric bilayer,
however; in fact there are no terms of this type allowed by the
symmetries of our problem.  In the absence of tilt
order all we can write down is the usual Canham-Helfrich free energy~\Can\Helfaa:
\eqn\eHshape{H_{\rm shape}=
\int\dd^2\xi\sqrt{g}\left[\mu_0+\half \,\ko(K_i{}^i)^2\right]~.  }
We have omitted the Gaussian curvature term because it is a total
derivative and we will not discuss topology change in this paper. In
particular as we will mention below total derivative terms do not
affect the renormalization of bulk terms.

We would like to emphasize a remarkable feature of this well-known formula:
it is invariant under 3-space parity transformations, even though we did
not insist on this! Indeed it is easy to write down chiral terms like
$\sqrt{g}\eps^{ij}\nabla_i\nabla_k K_j{}^k$, but every such term is
irrelevant (the term just quoted has length dimension $-1$). Thus, in fluid
membranes even if chirality is present in the constituent molecules, it
cannot be expressed on long scales; more precisely its effects on long
scales will be suppressed by powers of the scale difference.\foot{Something
similar happens in the standard model of particle physics: while we have
no reason to enforce baryon number conservation, we find that the other
symmetries of the theory forbid any relevant or marginal term violating this
symmetry. Thus, any $B$-violating processes must be suppressed by the scale
at which the standard model breaks down --- perhaps the grand unification
scale.} We will see how just the opposite situation obtains in the presence
of tilt order, giving a nice explanation to the observed relation between
chirality and tilt described in Section~2.

Turning now to include in-plane order, we must construct allowed low-dimension
terms involving the tilt order parameter $\hat m$.  
First we consider the pure tilt terms, \ie\ the terms which do not
contain the extrinsic curvature.  Since 
\eqn\twoeps{\eps_{ij}\eps_{kl}=g_{ik}g_{jl}-g_{il}g_{jk},}
we need only consider terms with no alternating tensor (the nonchiral
terms) or one alternating tensor (the chiral terms).  The lack of a
preferred normal \ztwo\ requires the chiral terms to have an odd number
of tilt fields and the nonchiral terms to have an even number.  By the
power counting described above, marginal terms have two covariant
derivatives while relevant terms have only one.  (Above we said that
covariant derivatives don't have any 3-space dimensions, and that is
correct.  However, since 2-space tensor indices end up getting
contracted with $g_{ij}$,
we can forget the distinction between 2-space
and 3-space dimensions for the purposes of listing all allowed
low-dimension terms.  In this way we immediately see that marginal
terms must have two covariant derivatives or two curvature tensors \etc)
Reparameterization
invariance and the fact that $\hat m$ is a unit vector lead to only
two marginal nonchiral terms:

\eqn\eHtilt{H_{\rm tilt}= {\ko\over2}
\int\dd^2\xi\sqrt{g}
[\gamma(\nabla_im^j)(\nabla^im_j)+\gamma'(\nabla\cdot\hat m)^2].}
In this formula we have written the tilt stiffnesses as
$\ko\gamma$, $\ko\gamma'$, since
generically (far from the tilt disordering transition)
$\gamma$, $\gamma'$ are pure numbers of order one.  
Coordinate indices are implicitly raised or lowered
using $g_{ij}$ and its inverse $g^{ij}$.   When we specialize to Monge gauge we will no longer
use this abbreviation; we will display all factors of $g_{ij}$ and
$e_\alpha{}^i$ explicitly.  The first term of \eHtilt\ has hexatic
symmetry; it is invariant under the shift $\theta\mapsto\theta +
{2\pi\over6}$ (in fact it is invariant under the
transformation $\theta\mapsto\theta + \alpha$ for constant $\alpha$).
This hexatic term leads to an increase of the effective bending stiffness,
counteracting the thermal softening that occurs in pure fluid
membranes \NePe\DaGuPe.  The second term of \eHtilt\ does not have
shift symmetry; it is the covariant generalization of  the anisotropic
term considered in~\NePelc.  

The only chiral
tilt term is a total derivative, the curl of $\hat m$:
$\eps^i{}_j\nabla_i m^j.$  This term is important for describing the
physics of defects \BrSe; it also plays a role in the mean
field theory description of tubules \OYL\Sel.
Once again, we will not consider total
derivatives here, so we drop this term; its effects are similar to
those of the term we retain~\HelPro. Another possible term is 
$\eps_{kl}m^i m^k \nabla_i m^l$, but it equals the curl term since
$\hat m$ is a unit 
vector.  This is most easily shown by writing the two expressions in
terms of $p^i=\eps^i{}_j m^j.$

There are several terms that involve both the tilt and the curvature.
These anisotropic terms signify a preference for the tilt order
parameter to align at some angle with a principal direction of
curvature.  The marginal terms have either two curvature tensors
or one curvature tensor and one covariant derivative.
Reparameterization invariance and the symmetry \ztwo\  again rule out any
marginal chiral terms.  For example, a marginal chiral term with two
curvature tensors would require an odd number of tilt fields; however,
an odd number of indices cannot be contracted to make an invariant.
The other case follows by similar reasoning.  The full list of
marginal terms is then
\eqnn\eHmixed
$$
\eqalignno{
H_{\rm mixed}=&
{\kappa_0\over2} \int\dd^2\xi\sqrt{g} \Bigl[
\alpha_1\hat m\cdot K\cdot K\cdot\hat m +
\alpha_2(\hat m\cdot K\cdot\hat m)(K_i{}^i) +
\alpha_3(\hat m\cdot K\cdot\hat m)^2\cr
&\qquad+\beta_1 (\hat m\cdot K\cdot\hat m)(\nabla\cdot\hat m) +
\beta_2(K_i{}^i)(\nabla\cdot \hat m)\cr
&\qquad+\beta_3 K^i{}_j\nabla_im^j+
\beta_4 m^\ell K_{\ell i}m^j\nabla_j m^i\Bigr]\quad. & \eHmixed}
$$
Again these formulas contain hidden metric factors;
for example $\hat m\cdot K
\cdot K\cdot\hat m = m^iK_i{}^jK_{jk}m^k$ where $K_i{}^j \equiv
K_{im}g^{mj}$.
Most of these terms have been considered before; \HelPro\ lists all of
these except the $K^i{}_j\nabla_im^j$ term, while \PePr\ is missing two
of the terms with one curvature tensor and one covariant derivative.

Turning to the possibility of chiral terms, there is
precisely one bulk term involving both the tilt field and the shape:
\eqn\eHstar{H_*=
{1\over 2} \cost\int\dd^2\xi\sqrt{g}\,m^i\eps_{ij}K^j{}_\ell m^\ell\;.}
This term has been discussed before~\DaLa; it enters along with the
$\eps^i{}_j\nabla_im^j$ term in Helfrich and Prost's
mean field theory of twisted ribbons \HelPro   \OYL\ as well as the 
mean field theory description of tubules~\Sel.  It also enters Peliti
and Prost's study of chiral smectic-${\rm C^*}$ membranes \PePr.

We note in passing that none of the terms in \eHmixed--\eHstar\ is
permitted by hexatic symmetry. Thus as asserted earlier, hexatic order
is no substitute for tilt order in expressing chirality at long scales.

The advantage of our 2d approach is this
exhaustive enumeration of all the allowed bulk terms.
We emphasize that we do not start with a 3d free energy and then
specialize to 2d, as other authors 
have \OYL\Sel.  For example, these authors begin
with the 3d Frank free energy for cholesteric liquid crystals and derive
both the $\eps^i{}_j\nabla_i m^j$ term and the $\hat m\cdot\eps\cdot K\cdot
K\cdot\hat m$ term from the spontaneous twist term.  In this approach
these chiral terms enter only in a specific linear combination.  In our
approach every independent invariant gets its own coupling; in
particular we expect any special relations among couplings not imposed
by symmetries to get spoiled under renormalization.  In practice
this is often a moot point since for convenience we either make a one
coupling constant approximation or simply truncate the free energy to
a few terms.  

As an aside we note that if the in-plane order is hexatic in character, then
most of the terms of \eHtilt\-\eHstar\ are forbidden, leaving 
only the kinetic term
studied by Nelson and Peliti~\NePe. In particular the chiral term
\eHstar\ is not allowed, and 
again we have an ``accidental'' parity symmetry. Only {\it tilt} can serve as the
vehicle to express chirality.\foot{Similarly in ref.~\BrSe\ 
 tilt is crucial for the {\it spontaneous} breaking of
parity.}

We note in passing that in a perturbative expansion about the zero-stiffness,
high-tension fixed point (Polyakov theory) many of the couplings in \eHshape
--\eHmixed\ are irrelevant and hence not usually written. However, the chiral
term \eHstar\ is marginal and could conceivably be one of interest, for
example in a random-surface theory of domain boundaries in a chiral 3d system.

\newsec{General Setup} 
\subsec{Measure}
From now on we will set Boltzmann's constant $k_B=1$.

To perform statistical sums we need to know how to count each configuration
just once. As mentioned, we can describe surfaces by the four independent
functions $(x^a(\xi),\hat m^i(\xi))$; $a=1,2,3$; $i=1,2$ (recall
$\|\hat m\|^2=1$), 
but this is a redundant description; replacing $\xi^i$ by $\xi^{\prime i}=
f^i(\xi)$ yields the same physical configuration. In the next
subsection we will fix this redundancy by choosing  a ``gauge'', but
first let us find an appropriate functional measure on the full
redundant configuration space. We will temporarily forget about the
angular variable $\hat m(\xi)$ since it poses no new conceptual issues once
$\vec x(\xi)$ is understood. Roughly speaking we want to construct a
measure as
\eqn\ePNma{\ppro\bigl[\pi^{-1/2}\dd x^a(\xi)\bigr] \quad,}
where the product is over all points of our surface, separated by a
cutoff distance. Eqn.~\ePNma\ needs a number of refinements before it
is correct, however.

Our problem is delicate because we work in a grand ensemble where the
number of elementary constituents (molecules or their representatives
after decimation) is not fixed; moreover the {\it density} of degrees
of freedom in $\xi$-space is not fixed but depends on $\vec x$ itself
via $g_{ij}=\pa_i\vx\cdot\pa_j\vx$. So \ePNma\ needs to be replaced by
some complicated measure nonlinear in $\vx$. We say some more about
this problem in Appendix A, but for our present purposes there is an
easy fix. We will work to one-loop order in fluctuations. This means
we approximate all functional integrals as {\it gaussian} integrals
about a chosen background configuration $\vxb(\xi)$ (see Section 4.3).
To do gaussian integrals we replace field space (the space of
$\vx(\xi)$) by its tangent space (the space of variations $\dl
\vx(\xi)$ about $\vxb(\xi)$) and approximate the energy
$H[\vx(\xi)]$ by a quadratic form in $\dl\vx(\xi)$. The measure
\eqn\ePNmb{\ppro\bigl[\pi^{-1/2}\dd\dl x^a(\xi)\bigr]}
now makes sense. Here the product is over points $\xi$ spread with
density $\Lambda^2\sqrt{\bar g}/\pi^2$, $\bar g=\det[\pa_i\vxb\cdot
\pa_j\vxb]$, and $\Lambda$ is our cutoff. The combination $\Lambda
g^{1/4}$ has no 3-space dimensionality, but $\dvx$ itself has
dimensions of length. To make our measure properly dimensionless we
therefore need a length scale; a convenient choice is the cutoff
$\Lambda$ itself. Allowing for an additional dimensionless factor
$\pi\eta $ at each point, we finally choose our measure to be
\eqn\ePNmc{[\dd\dvx]_\Lambda= \ppro\bigl[\sqrt\pi\eta\Lambda \dd
x^a(\xi)\bigr]\quad. }

Eqn.~\ePNmc\ is not yet in its most useful form; we would prefer a
continuum version. To get it we note that for a function $f(\xi)$ the
sum $\sum_\xi f(\xi)$ is approximately
$(\Lambda/\pi)^2\int\dgbxi f(\xi)$. Hence the measure \ePNmc\
can be equivalently specified by the requirement that 
\eqn\ePNmd{1=\int[\dd\dvx]_\Lambda\,\ex{-\Lambda^4\eta ^2\int \dgbxi\,(
\dvx)^2} \quad.}
In Appendix B we will choose $\eta^2 =\kappa_0/2T$ to make our formulas
simple, but other choices are possible.

It may seem disturbing that we have the freedom to choose $\eta $,
especially since we will see in Appendix B that different choices will
lead to different contributions to the renormalization of $\mu_0$.
Really, however, the measure is not a physical quantity, nor for that
matter are bare parameters like $\mu_0$, nor are quadratic divergences
universal. What's physical is the full
statistical weight $[\dd\vx]\ex{-H/T}$, which tells us how to compute
correlation functions. How we split this weight into $[\dd\vx]$ and
$\ex{-H/T}$ is our choice. As long as we choose $[\dd\vx]$ properly
local, as we have, we are assured that taking $H$ to be the most
general local functional we can describe any membrane with local
interactions. Two different measures will describe the same system by
two different sets of bare parameters; in particular the
scale-dependence of those parameters can appear different even though
each describes the same physics. 

One might think that at least $\eta$ should be chosen independent of all
parameters such as temperature, stiffness, \etc, so that \ePNmd\
defines a purely geometrical measure, but again this is not necessary,
nor even desirable. Consider for instance a much simpler system, an
$XY$ model in two dimensions. If we define our measure by requiring
$1=\int[\dd\varphi ]_\Lambda\,\ex{-\Lambda^2\int\dxi\,\varphi^2}$, then
the partition function $Z=\int[\dd\varphi]_\Lambda\,\ex{-{K\over T}
\int \dxi(\nabla\varphi)^2}$ has a constant term ${\Lambda^2
\over2 \pi^2}\log{T\over K}\int\dxi$ in $\log Z$. Interpreting this as a
renormalization of the constant term of $H[\varphi]$ (initially taken
to be zero), we find the latter to be nonanalytic in $T/K$! While ugly,
this result is not wrong; it just means that the naive choice of
measure has resulted in an inconvenient definition of the bare
constant energy term. Since constants drop out of correlation
functions we normally don't bother to cure this problem, but we could
easily do so by a new choice of measure setting $1=\int[\dd\varphi
]_\Lambda\,\ex{-{\Lambda^2K\over T}\int\dxi\,\varphi^2}$.

Similarly in membranes the choice \ePNmd\ will lead to inconvenient
quadratic divergences if we are not careful to choose $\eta $ properly.
These are more annoying than in ordinary field theory since
coordinate-invariance forces them to be not constants but proportional
to the surface area: they cause renormalizations of $\mu_0$.\foot{And
presumably of the Gaussian curvature coefficient as well.} 

\subsec{Gauge choice}

We next need to choose a ``gauge.''

We are interested in very stiff membranes. While the effective stiffness
decreases somewhat at long scales, it is still fairly large at the tubule
scale, as we see from the rigidity of tubules. Accordingly tubules are
fairly flat on the scale of the modes we wish to eliminate. A useful choice
of gauge is the ``Monge gauge,'' in which we choose $\xi^i$ on a given
surface so as to arrange that
\eqn\eMonge{\vec x(\xi)=(\ell\xi^1,\ell\xi^2,u(\xi^1,\xi^2))}
for some height function $u$. The constant $\ell$ has dimensions of 3-space
length; eventually we will adopt units where it equals unity. Some surfaces
cannot be represented in this form (those with overhangs), but at large
stiffness they are unimportant in the path integral. Physical questions, such
as the effective values of couplings in the Hamiltonian, are gauge-invariant;
we may calculate the answers in Monge gauge if that seems convenient, even if
we plan to use them to describe cylinders, for which Monge gauge certainly
does not work globally.

For our nearly-flat membranes $u$ will be small; eventually we will
expand our effective $H\eff$ in power series in $u$ and pick off our
renormalizations from the early terms of this expansion.

Note that the choice \eMonge\ is {\it inhomogeneous}; rescaling $\vec x(\xi)$
by a common constant spoils the two gauge conditions $x^i=\ell\xi^i$ $i=1,2$.
Other gauges such as conformal gauge ($(\partial_z\vec x)^2=0$ where $z=
\xi^1+i\xi^2$) are homogeneous; the distinction will be important when we
discuss field renormalization in Section~5.

To finish choosing a gauge, for each surface $u(\xi)$ we must choose an
orthonormal frame $\{\hat e_\alpha\}$, a set of vector fields obeying
$\hat e_\alpha\cdot\hat e_\beta\equiv g_{ij}e_\al{}^ie_\beta{}^j=\delta_{\alpha\beta}$.
In Monge gauge, for small $u$, we may choose
\eqn\eframe{e_\alpha{}^i=
\delta_\alpha{}^i - \half  \partial_\alpha u\partial^iu+\Order(u^4)\;.}
Here and henceforth we raise and lower indices using Kronecker deltas,
so that index placement does not imply hidden metric factors; similarly
we freely interchange index types. Thus, for example, \eframe\ says
$e_1{}^1=1-1/2(\partial u/\partial\xi^1)^2$. All metric and frame factors
will be shown explicitly.

Expressing $\hat m$ in terms of an angle field $\theta(\xi)$ using \emth\
and \eframe\ then gives us our desired nonredundant variables $(u(\xi),
\theta(\xi))$. To do statistical sums we now use the functional measure
\eqn\emub{[\dd\bar\mu]=\CJ[u]\cdot[\dd u][\dd\theta]}
where $\CJ$ is a Jacobian derived in Appendix A and the measures $[\dd u]$
and $[\dd\theta]$ are defined as in Section 4.1. Given the two nearby surfaces 
$\ub(\xi)$
and $\ub(\xi)+h(\xi)$, the invariant `distance' between them 
on function space is defined to be
\eqn\emetu{\|h\|_\ub^2\equiv\int\dd^2\xi\sqrt{\bar g}(h)^2}
where $\bar g_{ij}=\delta_{ij}+\partial_i \ub\partial_j \ub$ as usual. Similarly
we have that the `distance' between $\bar\theta$ and
$\bar\theta+\zeta$ is
\eqn\emettheta{\|\zeta\|_\ub^2\equiv\int\dd^2\xi\sqrt{\bar g}(\zeta)^2\;.}
Now $[\dd h]$, $[\dd\zeta]$ can be defined analogously to \ePNmd; for
example,
\eqn\emeau{1=\int[\dd h]_\Lambda\,\ex{-\Lambda^4\eta ^2 \|h\|_\ub^2} \quad.}

To finish specifying $[\dd\bar\mu]$ we need the Jacobian $\CJ[u]$. We compute
this in  Appendix A, but it is easy to see that we may neglect this factor
altogether for the calculation of Section~6. First, since \emetu, \emettheta,
and all other matrices involve $u$ only through the induced metric $g_{ij}$,
$\CJ[u]$ does not involve the extrinsic curvature and so cannot renormalize
terms like \eHstar\ directly.\foot{$\CJ$ {\it does} affect the area
term in \eHshape; see Appendix B.} Second, we will ultimately do a
saddle-point expansion in the 
stiffness $\ko/T$ appearing in $\eHshape$. The factor $\CJ$, being
purely geometrical in character, has no $\ko$ and hence Feynman-graph vertices
arising from it will, when inserted into a loop graph, add a
propagator and so increase its order
in $T/\ko$ by at least one power. Thus, the effects of $\CJ[u]$ on the
renormalization of the chiral coupling (and indeed that of $\kappa$ as well)
will only begin to appear at two loops.\foot{Polyakov makes a similar
observation in conformal gauge~\Pold.} Hence 
we will ignore $\CJ[u]$. Similarly different choices of the constant
$\eta $ in \ePNmd\ change our answers by the product over all points of
$\eta '/\eta $, which again is intrinsic and $\kappa_0$-independent and so
does not affect $\kappa$, $c^*$  to one-loop order.

\subsec{Effective action}

We would like to summarize the effects of fluctuations on short scales by
renormalization of the couplings. To one loop accuracy this is straightforward.
First recall how to summarize {\it all} fluctuations in an ``effective action''
$\Gamma[\ub]$. For any imposed $\ub$ we distort our system in such a
way as to ensure that the expected value of $u$ equals the given $\ub$.
We then calculate the logarithm of the partition function $Z$ for this
distorted system and call that ${-1\over T}\,\Gamma[\ub]$. For example,
if $\ub$ obeys the variational equations extremizing $H[u]$ then 
to one loop accuracy we can
simply distort the functional integral by imposing the boundary condition
$u(\xi)=\ub(\xi)$ on the edge of our surface; see \Polbook. More generally
we can introduce a source term $\int ju$ into $H[u]$ where $j(\xi)$ is a
2-form chosen to ensure $\langle u(\xi)\rangle =\ub(\xi)$ and let $\Gamma[\ub]=
{- T}\,\log Z[j]-\int j\ub$.

In practice these words are easy to carry out to one loop order. We simply
write
\eqn\eflu{u(\xi)=\ub (\xi) + h(\xi)}
(and later when we include tilt
\eqn\efltheta{\theta(\xi)=\bar\theta (\xi) + \zeta(\xi)}
as well) and substitute these into $H$. Then taking the source $j$ to be minus
the coefficient of the terms linear in $h$ (\ie, {\it dropping} such terms)
gives our $H'=H+\int ju$ an extremum at the desired $\ub$. Computing
with $H'$, in saddle-point approximation $\langle u\rangle$ is just the
extremum as desired. (To higher orders $H'$ is not quite what we want.)
Calculating our saddle-point integral now with $H'$, we may drop all terms
cubic and higher in $h$.

Accordingly let us define a quadratic form $Q_\ub[h]$ by
\eqn\eQf{H[\ub+h]=H[\ub]
+({\rm linear\; in}\; h)+
Q_\ub [h]+\Order(h^3)\quad.}
Thus $Q_\ub$ depends on the chosen background $\ub$. We wish to compute
\eqn\eeffa{
\Gamma[\ub]=H[\ub]-T\log\int[\dd u]\ex{-Q_\ub [u-\ub]/T}\;.
}
The functional integral can then be calculated in Gaussian
approximation, as follows. 

Abstractly, when we have a Gaussian integral of a quadratic form $Q(v)$ on a
vector space $V$,
$$
I=\int_V[\dd v]\ex{-Q(v)}
$$
we need to specify a measure $[\dd v]$ on a $V$ before we can assign an
answer to $I$. One way to do this is to stipulate that
\eqn\eabstr{1=\int_V[\dd v]\ex{-Q_0(v)}}
for some fixed reference $Q_0$. Then we simply 
have that $I=\det^{-1/2}\CO$,
where $\CO$ is the linear operator defined by\foot{Note that $Q$ itself has
no determinant; only the {\it ratio} $\CO$ of {\it two} quadratic forms has
a well-defined determinant.}
\eqn\eQQ{
Q(v)=(v,\CO v)_0\;,
}
the inner product corresponding to $Q_0$. In our problem we don't
really have a Gaussian integral, but we are making a Gaussian
approximation about $\ub$. Thus \emeau\ plays the role of \eabstr\
with $Q_0(h)=
\|h\|^2_\ub$.

Our formula \eeffa\ for the one-loop effective action thus
becomes\foot{In principle this 
formula could get modified by field renormalization. We show in
Section~5 that this does not happen.}
\eqn\efinalG{\Gamma[\ub]=H[\ub]+{T\over2}\log\det\CO_\ub\quad,}
where $\CO_\ub$ relates the quadratic bit $Q_\ub$ of
$H[u]$ near $\ub$ to the metric $Q_{0,\ub}=\|h\|^2_\ub$ (eqn. \emetu);
see \eQQ.  We have elaborated this
well-known formula because the metric is itself $\ub$-dependent; simply
attempting to take the determinant of $Q_\ub$ gives the wrong answer. This
$\ub$-dependence is the only vestige of the Liouville correction (see
Appendix~A) visible to one-loop order.

When we wish to summarize only certain modes our strategy is similar. We just
restrict our functional integrals, and ultimately the functional determinant
in \efinalG, to short-wavelength modes. In principle the correct way to do
this is to cut off invariantly using the 3-space distance $\Lambda^{-1}$,
\eg\ retain only modes whose eigenvalues for the covariant Laplacian
$\Delta=g^{ij}\nabla_i\nabla_j$ exceed $\Lambda^2$. Fortunately, however, to
our required accuracy we will see in Section~6 that a simpler
prescription suffices (in Appendix B this correction {\it will}
however be crucial). 

We will extend all the foregoing to include tilt. The validity of this
procedure may need some comment. By expanding $\theta(\xi)$ around a
background $\bar\theta$ (eventually taken to be zero below), we seem to
be assuming long-range order in the angular variable and doing spin-wave
approximation. But we know from the $XY$ model not to expect true
long-range order, and the spin-wave approximation can be misleading. Indeed
with stiffness $K$ we can compute the
two-point function by expanding in powers to $\theta$ to get
\eqn\exy{\langle\cos(\theta(x)-\theta(0))\rangle\propto 1-(T/2\pi K)\log|x|+
\ldots \quad,}
which seems to show long-range order, while actually the
low-temperature result is $|x|^{-T/2\pi
K}$. The point of course is that these expressions actually agree for
short lengths, and what we compute in an RG calculation is the effect
only of short-length modes. To get the correct answer we could either
sum an infinite set of terms, or compute the anomalous dimension for
$\cos\theta$ by a lowest-order expansion in $\theta$ and then solve
the RG equation, which effectively sums the same set of logs and
gives the same power-law behavior. Similarly the beta function we want for
$\cost$ can be accurately computed by an expansion in fluctuations of
$\theta$ about zero.

One may ask whether it is possible to compute the tilt part of our
functional determinant using the trick used in \DaGuPe\ for hexatic
order. In the latter case the only coupling of in-plane
order to shape is through the induced metric, and this led to
considerable simplification. In our case couplings like \eHmixed,
\eHstar\ couple tilt to extrinsic geometry and we must evaluate the
determinant explicitly.

\subsec{Some expansions}

We need to evaluate our bare Hamiltonian \eHshape, \eHtilt--\eHstar\ 
in Monge gauge,
then expand it about some background configuration $(\ub,\bar\theta)$ to get
its quadratic part $Q_{(\ub,\bar\theta)}[h,\zeta]$. Here $h$, $\zeta$ are 
the small fluctuations; see \eflu--\efltheta. Our answer \efinalG\ can be
expanded in powers of $\ub$, $\bar\theta$ about a flat, ordered surface with 
$\ub\equiv0$, $\bar\theta\equiv0$ (see the end of the previous subsection).
As we shall see we will only need the first few terms of this expansion.
Accordingly in this subsection we will carry out our calculation of
$Q_{(\ub,\bar\theta)}$ only to low orders in $\ub$, $\bar\theta$.

Eqns. \eHshape, \eHtilt--\eHstar\ contain a lot of terms. 
To make the problem manageable
we will severely truncate the theory by setting the couplings $\gamma'$,
$\alpha_j$ and $\beta_i$ all equal to zero. Certainly this is mathematically
consistent. All of the retained terms except $H_*$ enjoy greater
symmetry\foot{The symmetry sends $\theta(\xi)\mapsto\theta(\xi)+\alpha$ for
constant $\alpha$, {\it without} any corresponding change in the surface
shape. This is not just a 3-space rotation --- nor is there any physical
reason to impose it.} than the ones we have dropped, and so cannot induce
the latter upon renormalization. Moreover, since only $H_*$ breaks parity,
it too cannot induce the dropped terms to first order in its coupling $\cost$.
Since $\cost$ is small we will only compute to first order in it.

Physically this truncation to isotropic terms should give a reasonable
qualitative account of the effect of fluctuations on chirality. For
example, it is well known in the $XY$ model that the $\gamma'$ term
is less relevant than the retained $\gamma$ term \NePelc.

We now need a little differential geometry.  We set $\ell=1$ and
substitute the Monge gauge parameterization \eMonge\ into the formulas
for the various geometrical quantities defined in Section 3.  The
metric tensor is given exactly by 
\eqn\eMetric{g_{ij}=\delta_{ij}+u_i u_j,} where $u_i \equiv\del_i u$.
To the order required, the inverse metric tensor and the volume
element are
\eqnn\einvmet
\eqnn\evolelt
$$\eqalignno{g^{ij}=&\delta_{ij}-u_i u_j +
\Order(u^4)&\einvmet\cr\sqrt{g}=&1+{1\over2}u_i{}^2-{1\over8}u_i{}^2u_j{}^2+
\Order(u^6).&\evolelt\cr}$$ 
We will need the $\Order(u^4)$ part of the volume element in our
discussion of field renormalization.  As mentioned
earlier, repeated indices are summed using $\delta_{ij}.$  We have
already given the Monge gauge expression for our choice of the
zweibein $e_\alpha{}^i$ \eframe.  

We mentioned that given a physical configuration one of the ingredients
allowed in the Hamiltonian density is a covariant derivative. While
there are many such derivatives on a manifold, only one can be
specified using only the given geometrical data: the ``Levi-Civita''
connection, characterized by the properties that it has no torsion and
the metric tensor is covariantly constant. The latter condition
implies that the connection is given by a covector field $\Omega$:
\eqn\espin{\nabla_ie_\alpha\equiv\Omega_i\eps^\beta{}_\alpha e_\beta} where
$\ep^\beta{}_\alpha$ denotes the components of the alternating tensor
\ealter\ in an orthonormal basis (\ie\ $\eps^1{}_1=0, \eps^1{}_2=1$,
\etc). The first (no-torsion) condition says that 
$\nabla_i e_j - \nabla_j e_i = [e_i,e_j]$ where on the right we have
the Lie bracket. Substituting \eframe\ and \espin \ we can easily solve for $\Omega$:
\eqn\enotor{\Omega_i = {1\over 2} \eps^\alpha{}_\beta u_\alpha u_{\beta
i} + \Order(u^4).}
Finally, we can use the formulas \eframe, \espin, \enotor\ 
for the zweibein and spin connection
to compute the extrinsic curvature tension in an orthonormal basis:
\eqnn\eXcurv%
$$\eqalignno{K_{\alpha \beta}&=\hat n^a\cdot\nabla_\alpha\del_\beta
x^a\cr &=u_{\alpha \beta}-\half  u_{\alpha \beta}(u_{\gamma})^2 -
\half u_{\gamma}(u_{\alpha}u_{\gamma \beta}+ u_{\beta}u_{\gamma
\alpha}) + \Order(u^5).&\eXcurv\cr}$$
Using \eframe\ and \eMetric--\eXcurv, we find the truncated
Hamiltonian to be 
\eqnn\eHtrunc
$$\eqalignno{H&_{\rm
truncated}=\mu_0\int\dd^2\xi\bigl(1+{1\over2}u_i{}^2-
{1\over8}u_i{}^2u_j{}^2\bigr)\cr
&+{\ko\over2}\int\dd^2\xi\bigl((\Delta u)^2-{1\over2}u_i{}^2(\Delta
u)^2 -2(\Delta
u)u_iu_ju_{ij}\bigr)\cr
&+{\gamma\ko\over2}\int\dd^2\xi\bigl(\theta_i{}^2-
\theta_i\eps_{jk}u_ju_{ki}+{1\over2}u_i{}^2\theta_j{}^2-(u_i\theta_j)^2+
{1\over4}u_i{}^2(u_{kl})^2-{1\over4}(u_iu_{ij})^2\bigr)\cr
&+{\cost\over
2}\int \dd^2\xi\bigl(u_{12}+\theta(u_{22}-u_{11})+\theta
u_i(u_1u_{1i}-u_2u_{2i})\cr 
&\qquad\qquad-{2\over3}\theta^3(u_{22}-u_{11})-{1\over2}
u_j(u_1u_{2j}+u_2u_{1j}) - 2\theta^2u_{12}\bigr)\quad,&\eHtrunc}$$
where $\Delta=\del_i\del_i$ is the flat-space Laplacian, $\eps_{11}=0$,
$\eps_{12}= 1$ \etc, and $\theta_i\equiv\del_i\theta.$  We have kept
terms with at most four powers of the fields.

Our aim is to get the renormalized coefficient $c^*\eff$
of
${1\over2} \int\dd^2\xi\sqrt{\bar g}\,{\bar m}^i\eps_{ij}{\bar
K}^j{}_\ell {\bar m}^\ell\;,$  the chiral term in the effective action
\efinalG; $\ub, \bar m$ are the background fields.  Since we included
every  term that could arise except for total
derivatives, the truncated $H\eff$ must have the same form as $H_{\rm
truncated}$ with new values for the coefficients $\mu\eff$,
$\kappa\eff$ \etc\ (We show in the next section that there is no need
to rescale the field $\bar u$ to recover the original functional form
of $H_{\rm truncated}$, \ie\ there is no field
renormalization.)  For example, we can read off $\kappa\eff$ as the
coefficient of ${1\over2}(\Delta\bar u)^2$, the simplest chiral subterm in the
expansion of $\sqrt{\bar g}({\rm Tr}\bar K)^2$ (See \eHtrunc).  
Similarly we would like to read off $\ceff$ as the coefficient of the
simplest subterm $\half \ub_{12}$ in $H_{\rm eff}$, but now a problem
seems to arise. For one thing this subterm is a total derivative and
so vanishes with periodic boundary conditions whatever its
coefficient. We address this problem
by giving $\cost$ a fictitious spatial dependence for
intermediate stages of the calculation.  Thus, we will replace $\cost$
by $\cost f$, where $f=\cos(p\cdot\xi)$ and $p^i$ is a small wavenumber.
We will evaluate $H\eff$ on the particular
background configuration $\bar u = \bar u_0\cos(p\cdot\xi)$ so that
$\ceff$ emerges as the coefficient of $-\half p_1p_2\bar
u_0\cos^2(p\cdot\xi)$.
Even with this trick, our procedure may not correctly give us
$\ceff$ if there are neglected total derivative terms which also start off as $\bar
u_{12}$ in an expansion in $\bar u$ and $\bar\theta$.  In that case
only part of the coefficient of $\ubb$ will be $\ceff$; the rest will
be the effective coupling for the 
other, neglected, total derivative term.  However, the
only chiral total derivative term is $\eps^i{}_j\nabla_im^j$, which
involves only even powers of the height and so can not impersonate our
subterm $u_{12}$. 

Now that we know what we want to compute, the next step is to 
substitute $u=\bar u + h$, $\theta=\bar
\theta+\zeta$ into \eHtrunc\ to get the quadratic form $Q_{\bar u
\bar\theta}[h,\zeta]$ \eQf.  $Q$ will be of the form 
\eqn\eQofform{Q_{\bar u \bar
\theta}[h,\zeta]={\ko\over2}\int\dd^2\xi\sqrt{\bar g}\,(h,\zeta)\,\pmatrix{A&C\cr
D&B\cr}\, \pmatrix{h\cr\zeta\cr}}
where $\pmatrix{A&C\cr D&B\cr}=\CO_{\bar u \bar\theta}$ is a matrix of
differential operators depending on the background fields $\bar u$ and
$\bar\theta$.  We have pulled out a factor of $\sqrt{\bar g}$ in
accord with our definition of gaussian integrals \emetu\ and \eQQ.
The expressions for the elements of this matrix are quite lengthy;
since we are only interested in the renormalization of $\cost$ (and
$\mu_0$ in the next section) we will not write the terms involving
$\bar \theta$.  For convenience we will separate the terms into
$A=A_0+A_1+A_*,$ \etc\ where $A_0$ is the part of $A$ independent of $\ub,\
\bar\theta$, and the parameter $\cost$; $A_*$ is $\Order(\cost)$,
and $A_1$ contains the rest of the terms, which we carry only out to
second order in $\ub$. Even this is more than we'll need  --- we quote
these terms because they are useful in other calculations.
After some straightforward but lengthy algebra we find
\eqn\eoelts{\pmatrix{A_0&C_0\cr D_0&B_0\cr}=
\pmatrix{\Delta^2-{\mu_0\over\ko}\Delta&0\cr 0&-\gamma\Delta\cr}}
\eqnn\elong
$$
\eqalignno{A_1=&-{1\over2}\ub_i{}^2\Delta^2
+{1\over2}\del_\gamma(\Delta\ub)^2\del_\gamma -
{1\over2}\Delta(\ub_\gamma)^2\Delta +
2\del_\gamma\ub_\gamma(\Delta\ub)\Delta +
4\del_\alpha(\Delta\ub)\ub_\gamma\del_\gamma\del_\alpha\cr
&-2\Delta\ub_\gamma\ub_\alpha\del_\gamma\del_\alpha
+2\del_\gamma(\Delta\ub)\ub_{\gamma\alpha}\del_\alpha
+4\del_\gamma\ub_\alpha\ub_{\gamma\alpha}\Delta\cr
&+\gamma\Bigl[
-{1\over4}\del_\gamma(\ub_{\alpha\beta})^2\del_\gamma
+{1\over4}\del_\alpha\del_\beta(\ub_\gamma)^2\del_\alpha\del_\beta
-\del_\gamma\ub_\gamma\ub_{\alpha\beta}\del_\alpha\del_\beta
-{1\over4}\del_\alpha\del_\gamma\ub_\alpha\ub_\beta
\del_\beta\del_\gamma\cr
&
+{1\over4}\del_\alpha\ub_{\alpha\gamma}\ub_{\beta\gamma}\del_\beta
+{1\over2}\del_\beta\ub_\alpha\ub_{\alpha\gamma}\del_\beta\del_\gamma
+{1\over2}\del_\alpha\ub_\beta\ub_{\alpha\gamma}\del_\beta\del_\gamma\Bigr]
+A_{\mu_0}\cr
A_{\mu_0}=&{{\mu_0}\over\ko}\Bigl[{1\over2}\ub_i{}^2\Delta +
{1\over2}\del_\beta(\ub_\alpha)^2\del_\beta + \del_\alpha\ub_\alpha\ub_\beta\del_\beta\Bigr]\cr
B_1=&\gamma\Bigl[-{1\over2}\del_\alpha(\ub_\gamma)^2\del_\alpha
+\del_\alpha\ub_\alpha\ub_\beta\del_\beta+\half(\ub_\ga)^2\Delta
\Bigr]\cr
C_1=&{1\over2}\gamma\Bigl[-\del_\beta\del_\alpha\ub_\gamma
\eps_{\gamma\beta}\del_\alpha
+\del_\gamma\ub_{\beta\alpha}\eps_{\gamma\beta}\del_\alpha\Bigr]\cr
D_1=&{1\over2}\gamma\Bigl[\del_\alpha\eps_{\gamma\beta}
\ub_\gamma\del_\beta\del_\alpha 
+ \del_\alpha\ub_{\beta\alpha}\eps_{\gamma\beta}\del_\gamma\Bigr]&\elong\cr
}$$
Similarly carrying the chiral terms out only to first order in $\ub$,
\eqnn\elongb
$$
\eqalignno{
A_*=&{\cost\over2\ko}\Bigl[\del_1 f\ub_\gamma\del_2\del_\gamma
+\del_\gamma f\ub_1\del_2\del_\gamma
+\del_1f\ub_{2\gamma}\del_\gamma +(1\leftrightarrow2)\Bigr]\cr
C_*=&{\cost\over{2\ko}}\Bigl[\del_2{}^2f-\del_1{}^2f\Bigr]\cr
D_*=&{\cost\over{2\ko}}f\Bigl[\del_2{}^2-\del_1{}^2\Bigr]\cr
B_*=&-{2\cost\over{\ko}}f\ub_{12}\quad.&\elongb\cr
}
$$
In these formulas the
derivatives not in parentheses act on everything to the right.  Care
must be taken when integrating by parts to get these 
formulas since $f$ can get differentiated.  All that remains is to
expand the logarithm of $\det\Order_{\ub \bar\theta}$ to first order in
$\cost$ to get $\ceff$.  But first we digress to discuss
field renormalization and the area term. 

\newsec{Field Renormalization and the Area Term}
\subsec{A Ward identity}
The Hamiltonian \eHshape, \eHtilt--\eHstar\ is the most general local
expression of low dimension operators respecting the symmetries of a
lipid bilayer.  If this Hamiltonian is expanded in a powers series in
the fields $u$ and $\theta$, certain coefficients of this expansion
will be related by the symmetry; \eg\ the coefficient of
$u_i{}^2u_j{}^2$ is $-{1\over4}$ the coefficient of $u_i{}^2$ (see
\eHtrunc).  The coefficients in the expansion of $H\eff$ in $\ub$ and
$\bar \theta$ likewise must be related, since the measure and the
cutoff both respect all the symmetries.  However, it is possible that
a rescaling of fields may be necessary to recover the original
relationships among the coefficients, \ie\ there may be a need for
field renormalization, as in the usual nonlinear sigma model~\NePelc%
.  In this section we prove that
rotational invariance implies that there is no field
renormalization of $u$.  While this fact is known
\PeLe\ForGab\WaZia, we think it deserves some discussion.

Intuitively, it is clear that rotational invariance prevents field
renormalization in our choice of gauge: if the height field
$x^3(\xi)=u(\xi)$ gets rescaled, then by rotational invariance the
other two coordinates $x^{1,2}$ must get rescaled as well.  But this
rescaling would spoil the inhomogeneous Monge gauge condition \eMonge.
Let us make this intuition more precise. 

We will give a simple nonperturbative argument that proves that there
is no field renormalization of the height field $u$
in Monge gauge, even in the
presence of {\it all} the couplings of \eHshape, \eHtilt--\eHstar. (We
will not need to investigate the angle field.) This argument 
closely follows the reasoning in \LTR\ref\GoLu{L. Golubovi\'c and T.
C. Lubensky, ``Smectic elastic constants of lamellar fluid membrane
phases: Crumpling effects,'' Phys. Rev. {\bf B39} (1989) 12110.}.  We
will consider a square frame of physical
dimensions $L\times L$ and impose the boundary condition of vanishing
height $u$ at the boundary

For simplicity we will suppress the tilt fields from the following formulas.
The original Hamiltonian $H[\vec x(\xi)]$ is rotationally invariant.
Thus casting it into Monge gauge the functional form of $H[u(\xi)]$
must be independent of 
{\it which} Monge gauge we choose, \ie\ which plane in space we choose
to measure 
the height $u$. The functional measure and cutoff are likewise
rotationally invariant, and so the effective Hamiltonian $H_{\rm eff}$
must also have this property. 

We will examine the same physical surface in two different Monge
gauges, leading to two different height fields $u$, $\tilde u$. Requiring
 $H\eff[u]=H\eff[\tilde u]$
will give certain
relationships among the various terms of \eHshape, \eHtilt--\eHstar\ in an
expansion in $u$.  In particular, we will see that these
relationships are identical to those in the original Hamiltonian, $H$, so
no field rescaling is needed.

Let us consider a surface with cylindrical symmetry,\foot{There is no
loss of generality here. Terms in the effective action involving
$\pa_2 u$ are related to the ones retained here by in-plane rotation
invariance. }
 so that
$u(\xi^i)= u(\xi^1)$, and let the new plane defining the new Monge
gauge be the old plane, rotated about the $\xi^2$-axis by an angle
$\psi$
(see \tfig\fsideview).
{\def\y{\xi^1}\def\ut{\tilde u}\def\yp{\xi^{\prime1}}\def\cp{\cos\psi}
 \def\x{\xi^2}\def\xp{\xi^{\prime2}}
Then $u(\y)$ is the perpendicular distance from $P$ to the original
plane, while $\ut(\yp)$ is the perpendicular distance from $P'$ to the
rotated plane. Thus
\eqn\ePNa{\yp=\y \cp\,\qquad \ut(\yp)=\y\sin\psi+\dl\quad,}
where
$$\dl\cp=u(\y+\dl\sin\psi)\quad.$$
Solving, we find that
\eqn\ePNb{\dl(\y)=u(\y)+\psi u(\y)u'(\y)+\psi^2[\half u(\y)+u'(\y)^2u(\y)
+\half u^{\prime\prime}(\y)u(\y)^2]+ \Order(\psi^3)\quad,}
where $u'\equiv \pd u{\y}$ \etc

\ifigure\fsideview{Side view of membrane and rotated reference axes.
}{fsideview.eps}{1.6}

Rotation invariance now says that $H[u(\y)]=H[\ut(\yp)]$. Since $H$ is
local we thus have 
$$\int_0^L \dd\y\int_0^L \dd\x \CH[u(\y,\x)] =
  \int_0^{L\cp} \dd\yp\int_0^L \dd\xp \CH[u(\yp,\xp)]$$
for some universal functional $\CH$ (here $\xp=\x$). Expanding $\CH$
in derivatives this says that
\eqn\ePNc{\eqalign{\int_0^L \dd\y\ \CH(u,u',\ldots) =&
\int_0^{L\cp}\dd\yp\ \CH\Bigl(\ut(\yp),\pd{\ut}{\yp}\Bigr)\cr
=& \int_0^L(\cp\dd\y)\,\CH\Bigl(\ut(\y),{1\over\cp}\ut'(\y)\Bigr)\quad.\cr}}
Let us compare on both sides the terms which can be brought to the
form $(u')^n$ by parts integration. Such terms can only arise from the
area term of \eHshape \ (and similarly when we introduce tilt); call
them $$\CH=\mu_0+\mu_2(u')^2+\mu_4(u')^4+\cdots\quad.$$
Thus the ratio of any two $\mu_n$ will tell us about field
renormalization, since in the original $H$ we have $\mu_2=\half\mu_0$, \etc{}
Substituting \ePNb\ into \ePNa\ and then into \ePNc, we at once find
at $\Order(\psi)$ that $H\eff$ has
$\mu_4=-\lfr14 \mu_2$, and at $\Order(\psi^2)$ that 
$\mu_2=\half\mu_0$. Thus there is no field renormalization, as claimed.

}

 We could continue in this way to
find the other $\mu$-type coefficients; what we would find is that the
terms with as many derivatives as height fields enter the effective
Hamiltonian only through the invariant combination 
$$\mu_0\int\dd^2\xi\sqrt{1+\ub_i{}^2}.$$

\subsec{Other gauges}

The absence of field renormalization in Monge gauge is just a
manifestation of a more general idea: physical 3-space distances do
not scale anomalously under the appropriate renormalization group
transformation, since it too is a scaling of 3-space distances. In any
gauge the fields $\vec x(\xi)$ take values in physical space, so when
we scale distances by a factor of $b$, $\vec x\mapsto b\vec x$. The
coordinates $\xi$ themselves live in an unphysical parameter space and
so don't have any {\it a priori} transformation law. Thus the
situation is just the reverse of the usual nonlinear sigma model: here
the spin field ${\bf s}(\xi)$  takes values in some internal space and
its scaling is not so obvious, while $\xi^i$ have values in ordinary
space. So we would expect to find no field renormalization in any
gauge (apart from a canonical dimension if we choose to rescale
everything to restore the original cutoff). Let us digress slightly to
make this intuition more precise.

Monge gauge is an inhomogeneous gauge in the sense that \eMonge\
contains a dimensionful parameter $\ell$; thus a single term of the
covariant $H$ \eHshape\ will give rise to terms of varying degree in
$u$, even though each term of \eHshape\ is homogeneous in $\vec x$.
 (We obscured this by  choosing units with $\ell=1$ in \eHtrunc\ and
elsewhere.) Hence we cannot rescale $u$ at will; simply calculating
$H\eff[\ub]$ and inspecting different subterms of the area term
suffices to determine what rescaling of $\ub$ if any is required. In
Appendix B we sketch such an explicit calculation
, but the previous subsection arrived at a more general
conclusion by exploiting a symmetry transformation \ePNa--\ePNb\ which
was also inhomogeneous in $u$.

Other popular gauge conditions are homogeneous, however. For example,
normal gauge requires 
\eqn\enorga{(\vec x-\vec{\bar x})\cdot\del_i\vec{\bar x}=0\qquad i=1,2}
where $\vec{\bar x}$ is an arbitrary reference surface. Eqn.~\enorga\
contains no constant like $\ell$, and indeed it is well known that
after one-loop integration of the fast modes $H\eff[\vec{\bar x}\,]$ has
the same functional form as the original $H$ \DaLe, and in particular
retains its form under rescaling of $\vec{\bar x}$. Since such a
rescaling will affect the effective values of dimensionful couplings
like $\mu\eff$, $\cost$, we need some other means to determine how
much if any rescaling of $\vec x$ is needed.

The idea of renormalization is that if one cut-off field theory
represents the continuum limit of discrete constituents of some size
$\Lambda\inv$, then we can find another equivalent field theory
representing new, effective constituents of physical size
$(b\Lambda)\inv >\Lambda\inv$. For the latter (decimated) theory to
reproduce the long-scale results of the former one, all coupling
constants corresponding to intrinsic properties of the molecules must
be given new renormalized values. The new values are not known {\it a
priori}; we must solve for them. So these terms won't help us find out
how the field rescales. Instead we need to consider the coupling of
some {\it external} thermodynamic force to the system, one whose
meaning is clear both for the original contituents and for the larger
effective ones.

Looking at \eHshape, at first it seems that the area coefficient
$\mu_0$ is an appropriate choice, as it is usually interpreted as an
applied chemical potential for a reservoir of amphiphiles; if the
effective constituents are $b\inv$ times as big as the original ones,
their free energy cost should also be $b\inv$ as great. But really
$\mu_0$ is a free energy cost per molecule {\it divided by} the area
per molecule, and it is not clear how the latter quantity, intrinsic
to the molecules, behaves under decimation. After all the decimated
surface is the best approximation to the original one made with the
coarser molecules; it will have smaller area if the original surface
is rough on the scale of the cutoff. So indeed we expect $\mu$ to
suffer nontrivial renormalization, which we need to compute. Thus the
area term cannot be used to fix the rescaling of $\vec x$.

Fortunately another, independent, thermodynamic force term is
available which {\it is} purely extrinsic to the constituents: We can
confine our membrane to a frame of physical size $L$ via a boundary
condition like $x^1(0,\xi^2)=0$, $x^1(1,\xi^2)=L$, \etc{} Of course
$L$ means the same thing for both the original constituents and the
larger effective ones: each must make a net spanning the same physical
frame. So we must not rescale $\vec x$ as part of our RG
transformation (except possibly for the trivial canonical rescaling
mentioned above). This is in fact the formulation used for example by
David and Leibler (their Eqn.~21) \DaLe\ in normal gauge, and
Polyakov~\Pold\ in conformal gauge\foot{In \Pold\ Polyakov does
not really rescale the fields $x$, rather the rescaling there amounts
to integrating out the Lagrange multiplier field. }.

The above remarks just amount to an elaboration of the argument at the
beginning of the subsection: $\vec x$ really lives in physical space,
where external probes can literally measure its length. So we have no
freedom to give $\vec x$ an anomalous dimension.

\subsec{$\mu_0$ is $\Order(T)$}
We argued above that the area coefficient $\mu_0$ will get
nontrivially renormalized. We now want to argue that for our purposes
it suffices to neglect the area term completely, \ie\ we may take
$\mu_0=0$ altogether. At first this seems to contradict our
`naturalness' argument, which by dimension counting gives $\mu_0$ a
natural value of order an electron volt times $\Lambda^2$. To see why
this expectation is wrong, we again begin with an intuitive
discussion.

Physically we work in a fixed-area ensemble. A certain number of
amphiphiles are put in water and form surfaces of total area $A$
proportional to the fixed number of molecules. We may find it more
convenient to work instead in an ensemble of fixed area cost $\mu_0$,
later performing a  Legendre transform back to fixed $A$ by choosing
$\mu_0$ so that surfaces of area $A$ dominate the statistical sum.
Now examine the free
energy $H_{\rm shape}$. On one hand, the area term wants the surface
area to shrink without limit; it is happiest at zero size. On the
other hand, we don't want to omit this term altogether, as then
surfaces of all possible areas would enter the statistical sum due to
the scale invariance of the curvature term. Clearly what we want is to
choose a value of $\mu_0$ on the order of  $TA\inv$, so that
this term begins to hurt for surfaces of size $A$. For large $A$ this
value of $\mu_0$ is negligible compared to the `natural' size $\Lambda^2$. 

The above argument neglects the effects of fluctuations. If we set
$\mu_0$ to zero our system may nevertheless induce an effective area
term due to the fact that highly corrugated surfaces have a lot of
entropy. Indeed David and Leibler have shown that the appropriate
condition for freely-floating surfaces is not $\mu_0=0$ but rather
that the renormalized area coefficient evaluated at the size of the
system be zero~\DaLe\ (or rather $A\inv$, which is tiny compared to
$\Lambda^2$). This 
$\mu\eff$ is the area coefficient of the full effective Hamiltonian
with all fluctuations integrated out; setting it close to zero
therefore allows very large surfaces to dominate the full statistical
sum, as desired. If
$\sqrt A$ exceeds the persistence length $\xi_p$
of the system then corrugations are very important and this condition
may indeed require a large value of $\mu_0\propto\Lambda^2$. If this
happens then our entire perturbation theory breaks down since $\xi_p$
is the scale where $T/\ko\sim1$. The path integral is then dominated
by spiky configurations and is not truly two-dimensional at all
\ref\Cates{M. Cates, Phys. Lett. {\bf 161B} (1985) 363.}.

In our
systems, however, we know experimentally that shape fluctuations are
not yet large on 
the scale of our structures.\foot{In fact for a nonchiral membrane
with in-plane 
order the persistence length is {\it infinite} due to a line of fixed
points at weak coupling and low tension~\NePe. Technically
this is our real point of departure from string theory, which assumes
\Pold\Polbook\ that there is no fixed point, so that we flow to strong
coupling (a fixed point of zero stiffness and high tension). Chirality
destabilizes this line of 
fixed points, but at weak chirality we still expect to stay close to
a weakly-coupled fixed point.}
 Then we expect that the required value of
$\mu_0$, while not zero, will be suppressed by one power of
temperature.    In Appendix B we review the calculation of
$\mu\eff$ to one loop and indeed find that 
\eqn\emueff{\mu\eff(b^{-1}\Lambda^{-1})=\mu_0-T{\Lambda^2\over
\pi^2}\log b^{-1}}
for $b$ close to unity, plus less-divergent terms.
The bare value of the area term must therefore be of the form
$\mu_0=\bar\mu_0 T\Lambda^2$, where $\bar\mu_0$ is a dimensionless
number of order unity.

\newsec{Renormalization of the Chiral Coupling}

We finally turn to the calculation of the effective chiral coupling
$\ceff(b^{-1})$, which summarizes the effect on chirality of fluctuations with
wavevectors between $b\Lambda$ and $\Lambda$.  

As mentioned in Section
3.3, there is no need to include total derivative terms in this
calculation since these 
terms cannot induce bulk terms like $\cost$. One can see this by
noting that 
since we are eliminating only short-wavelength modes,
we can choose a basis for these modes
consisting of wavepackets with support only in the bulk.
In this basis it is clear that boundary terms cannot lead to
fluctuation-induced bulk terms, since the boundary terms are strictly
zero.  Bulk terms {\it can} however lead to fluctuation-induced boundary
terms; that is why we had to worry about induced chiral boundary terms
mimicking our bulk boundary term in Section~4.4.

We claimed in Section~5.3 that the area coefficient $\mu_0$ will not
contribute to the one 
loop result for $\ceff$, since $\mu_0$, being induced by fluctuations,
is already $\Order(T)$.  We will see this explicitly in our calculation.
Also, we argued in Section 2 that since the chirality of  typical membrane
lipid molecules is a rather minor aspect of their shape, we expect
$\cost$ to be 
small compared to the cutoff.  Our task is therefore to expand
${T\over2}\log\det\CO_{\ub\bar\theta}$, the operator appearing below
\eQofform, to first order in $\cost$. More precisely, we want the
contribution to this determinant from modes between the cutoff
$\Lambda$ and a slightly smaller cutoff $b\Lambda$. We review the
meaning of such determinants in Appendix~B. As discussed in Section~4.3,
we only need to carry $H\eff$ out to $\Order(\ub)$.

We will
not write the $\cost$-independent terms.  The desired terms of the
expansion are therefore
\eqnn\eXpandlog
$$\eqalignno{
{T\over2}{\rm Tr}&\log 
\pmatrix{A_0& C_0\cr D_0& B_0}\biggl[{\spec I}
+\pmatrix{A_0& C_0\cr D_0& B_0}^{-1} \pmatrix{A_1+A_*& C_1+C_*\cr
D_1+D_*& B_1+B_*}\biggr]\cr
=&{\rm const}+{T\over2}{\rm
Tr}\pmatrix{(\Delta^2-{\mu_0\over\ko}\Delta)^{-1}& 0\cr 0&
-\gamma^{-1}\Delta^{-1}}\pmatrix{A_*& C_*\cr D_*& B_*}\cr
&+{T\over2}\left({-\half}\right)(2){\rm
Tr}\pmatrix{(\Delta^2-{\mu_0\over\ko}\Delta)^{-1}& 0\cr 0&
-\gamma^{-1}\Delta^{-1}} \pmatrix{A_*& C_*\cr D_*& B*}\cr
&\times\pmatrix{(\Delta^2-{\mu_0\over\ko}\Delta)^{-1}& 0\cr 0&
-\gamma^{-1}\Delta^{-1}} \pmatrix{A_1& C_1\cr D_1& B_1}\quad.&\eXpandlog}
$$
Here $A_0$ \etc\ are defined in \eoelts, $A_1$ \etc\ in \elong, and
$A_*$ \etc\ in \elongb. In the first term on the r.h.s.\ we  need
$A_*$ \etc\ to $\Order(\ub)$. This term can thus
be represented by the Feynman graphs of \tfig\fcstgra.
\ifigure\fcstgra{Feynman graphs representing the first term in the
expansion of the logarithm.}{fcstgra.eps}{1}
\noindent In the second term we keep $A_*$ \etc\ to zeroth order in $\ub$ since
$A_1$ \etc\ are already $\Order(\ub)$. This gives a graph with two
propagators, \tfig\fcstgrb.  \ifigure\fcstgrb{Feynman graph representing
the second term in the expansion of the logarithm.}{fcstgrb.eps}{1.3}
Consider first  \fcstgra:
\eqnn\egra
$$
\eqalignno{{T\over2}{\rm Tr}&\pmatrix{
\bigl(\Delta^2-{\mu_0\over\ko}\Delta\bigr)^{-1}& \cr
&-\ga\inv\Delta\inv\cr}
\pmatrix{A_*\vphantom{\bigl(\Delta^2-{\mu_0\over\ko}\Delta\bigr)^{-1}}&C_*\cr
 D_*\vphantom{\ga\inv\Delta\inv}&B_*\cr}\cr
&={\cost\over\ko} {T\over2}{\rm
Tr}\Biggl\{(\Delta^2-{\mu_0\over\ko}\Delta)^{-1}\cr
&\qquad\times\Bigl[\half\bigl( \del_1(f\ub_\gamma)
+ \del_\gamma(f\ub_1)\bigr)\del_2\del_\gamma
+\half f\ub_{2\gamma}\del_1\del_\gamma +(1\leftrightarrow
2)\Bigr]\cr
&\qquad+2\gamma^{-1}\Delta^{-1}f\ubb\Biggr\}\quad .&\egra}
$$
The first two terms of \egra\ are truly
total derivatives, and may be dropped. We have also used rotation
invariance to drop terms with an odd number of derivatives not
differentiating $\ub$ or $f$. Rotation invariance also lets us replace 
$\del_1\del_\gamma$ by ${1\over2}\delta_{1\gamma}\Delta$.  

In \egra\ we should use the eigenvalues of the covariant Laplacian to
regulate the trace. As described in Appendix B, these can be
related to the eigenvalues of
the flat Laplacian using perturbation theory.  Since the height $\ub$
enters the covariant Laplacian through the metric $\bar g_{ij}$, the 
errors we make here are of $\Order(\ub^3)$ and do not affect our
result. Hence \egra\  becomes
\eqn\egraII{\egra =
{\cost\over\ko}{T\over2}\int\dd^2\xi\bigl[(\half+2\gamma^{-1})\ubb
f\bigr]\,\int^\Lambda_{b\Lambda}
{{\rm d^2}k\over(2\pi)^2}
{-1\over{k^2+{\mu_0\over\ko}}}+\Order(\ub^3).}
If we write $\mu_0={\bar\mu_0}T\Lambda^2$ as in Section 5.3 and note
that $b$ is close 
to unity (\ie\ work to first order in $\log b^{-1}$), then the
$k$-integral is easily computed to give
\eqn\egraIII{\egra =-\cost{T\over{4\pi\ko}}{\log
b^{-1}\over{1+{{\bar\mu_0}\over\ko}}T}\int\dd^2\xi(\half
+2\gamma^{-1})\ubb f.}  
Note that the divergence of \egraIII\ is logarithmic, not linear as
might be expected by a naive application of the power-counting rules of
Section 3.  Since the coupling $\cost$ is the only parity-violating
coupling in the theory, its one loop correction must be proportional
to $\cost$ itself.  Dimensional analysis therefore implies that the
divergence of each of the one loop graphs of \fcstgra--\fcstgrb\ is at most
logarithmic.\foot{This situation is analogous to the renormalization
of the mass term in quantum electrodynamics:  the mass term
$m_0\bar\psi\psi$  is the only term that 
breaks the $U(1)$ chiral symmetry; therefore, its one loop correction
is proportional to $m_0$ itself and so is
logarithmic, not linear, in the cutoff.} 

Returning to \egraIII, we see as promised that the $\mu_0$ term gives a
two loop contribution and so a small effect as long as
$\mo/\Lambda^2\ko\ll 1$. This holds as long as $\ko\gg T$, \ie\ at
length scales below the persistence length, if any.

We now turn to the graphs represented by \fcstgrb.
Again, we only write the terms that induce the $\ubb$ term.  
The second term of \eXpandlog\ contributes
\eqnn\elogII
$$\eqalignno{-&{T\over2}{\rm Tr}\pmatrix{0&
(\Delta^2-{\mu_0\over\ko}\Delta)^{-1}C_*\cr
-\gamma^{-1}\Delta^{-1}D_*& 0}\pmatrix{0&
(\Delta^2-{\mu_0\over\ko}\Delta)^{-1}C_1\cr
-\gamma^{-1}\Delta^{-1}D_1& 0}\cr
&={T\over2}\gamma^{-1}{\rm
Tr}\bigl[(\Delta^2-{\mu_0\over\ko}\Delta)^{-1}C_* \Delta^{-1}D_1
+\Delta^{-1}D_*(\Delta^2-{\mu_0\over\ko}\Delta)^{-1}C_1\bigr]\cr
&={T\over2}{\cost\over\ko}{1\over4}{\rm
Tr}\bigl((\Delta^2-{\mu_0\over\ko}\Delta)^{-1}(\del_2{}^2
-\del_1{}^2)f\Delta^{-1}
(\eps_{\gamma\beta}\ub_\gamma\del_\beta\Delta)\cr
&+\Delta^{-1}f(\del_2{}^2-\del_1{}^2)
(\Delta^2-{\mu_0\over\ko}\Delta)^{-1} \ep_{\ga\beta}
(-2\ub_{\alpha\gamma}\del_\beta
-\ub_\gamma\del_\beta\del_\alpha)\del_\alpha\bigr)\quad,&\elogII}$$
where we have retained only those terms with two or fewer derivatives
on $\ub$ or $f$ (other terms are not divergent and do not renormalize
$\cost$), and we used the antisymmetry of $\ep_{\ga\beta}$ to
eliminate some terms.  Now we will evaluate these traces using the
particular forms for $\ub$ and $f$ chosen in Section 4.  Passing to
complex notation, we use
\eqnn\euandf
$$\eqalignno{f=&f_0\ex{-ip\cdot\xi}\cr\ub=&\ub_0
\ex{ip\cdot\xi}\quad.&\euandf}$$
The functional traces are then
\eqnn\etraces
$$\eqalignno{\elogII
=&\ub_0{T\over8}{\cost\over\ko}\int_{b\Lambda }^\Lambda{{\rm
d^2}k\over(2\pi)^2}k^2k_\beta
p_\gamma\eps_{\gamma\beta}{1\over(k+p)^2}(k_2{}^2-k_1{}^2)
{1\over k^4+{\mu_0\over\ko}k^2}\int\dd^2\xi f_0\cr
+&\ub_0{T\over8}{\cost\over\ko}\int_{b\Lambda }^\Lambda {{\rm
d^2}k\over(2\pi)^2} k_\alpha\eps_{\gamma\beta}(-2k_\beta p_\alpha p_\gamma
-k_\beta k_\alpha
p_\gamma)\cr &\qquad\times{1\over(k+p)^4+{\mu_0\over\ko}(k+p)^2}((k_2+p_2)^2
-(k_1+p_1)^2) {1\over k^2}\int\dd^2\xi f_0\ .&\etraces}
$$
It's convenient to shift the integration variable in the second
integral of \etraces\ $k+p\mapsto k$.  We do not shift the
range of integration; the error incurred is of the form 
\eqn\eerror{\int_{b\Lambda}^{ \Lambda}[F(k)-F(k-p)]{\rm
d^2}k=\int_{b\Lambda }^{ \Lambda}p\cdot{\del F\over\del
k}{\rm d^2}k+\Order(p^2).}
Since the integrand $F$ of the second integral in \etraces\ is of the
form $g(p)/k^2 +\Order(1/k^3)$ ($g$ is some function of $p$), the difference
\eerror\ will be of order $1/\Lambda$.  The second integral in
\etraces\ is thus
\eqnn\esecint
$$\eqalignno{\ub_0{T\over8}{\cost\over\ko}&\int_{b\Lambda}
^{ \Lambda}{{\rm
d^2}k\over(2\pi)^2} \eps_{\gamma\beta}(-k_\alpha k_\beta k_\alpha
p_\gamma)\cr &\times{1\over k^4+{\mu_0\over\ko}k^2}
(k_2{}^2-k_1{}^2){1\over(k-p)^2}\int\dd^2\xi f_0&\esecint}$$
where once again terms of $\Order(\Lambda^{-1})$ have been dropped.
Expanding 
$${1\over(k-p)^2}={1\over k^2} +2{k\cdot p\over k^4} +\Order(k^{-4})$$ 
and using
$$\eqalignno{\int_{b\Lambda }^{ \Lambda}{{\rm
d^2}k\over(2\pi)^2}{k_\alpha k_\beta k_\eps k_\lambda\over
k^2(k^4 +{\mu_0\over\ko}k^2)}={1\over8}\int_{b\Lambda}^{\Lambda}&{{\rm
d^2}k\over(2\pi)^2}{1\over
k^2+{\mu_0\over\ko}}\cr&\times(\delta_{\alpha\beta}\delta_{\eps\lambda}+
\delta_{\alpha\eps}\delta_{\beta\lambda}+
\delta_{\alpha\lambda}\delta_{\beta\eps})} $$
we find that \esecint\ is simply 
$$-\ub_0{T\over8}{\cost\over\ko}\int_{b\Lambda}^{\Lambda}{{\rm
d^2}k\over(2\pi)^2}{1\over k^2+{\mu_0\over\ko}} p_1p_2\int\dd^2\xi f_0$$
or (see \euandf)
\eqn\esecintII{\esecint={T\over4\pi\ko}{\cost\over4}{\log b^{-1}\over
1+{{\bar \mu_0}\over\ko}T} \int\dd^2\xi f\ubb.}
The first integral of \etraces\ can be evaluated in a similar way to
give a contribution of 
\eqn\efirint{{T\over4\pi\ko}{\cost\over 4}{\log b^{-1}\over
1+{{\bar \mu_0}\over\ko}T} \int\dd^2\xi f\ubb.}
Once again we see that the $\mu_0$ counterterm leads to a two loop
contribution; we will ignore it henceforth.
The total contribution to $H_{{\rm eff},*}$, the chiral part of the one
loop effective Hamiltonian, from \egraIII, \esecintII, and \efirint\ is 
\eqn\echiham{H_{{\rm eff},*}={\cost\over
2}\int\dd^2\xi(1-{T\over\pi\ko\gamma}\log b^{-1})\ubb.}
Our final result is that the effect of fluctuations on chirality may
be summarized by omitting them but replacing $\cost$ by $\ceff$, where
\eqn\ebtafcn{{{\rm d}\ceff(b^{-1})\over{\rm d}\log b^{-1}}=-{\ceff
T\over\pi\ko\gamma}.}
We make no errors to one loop in \ebtafcn\ if we replace $\ko$ and $\gamma$ by
their effective values; note that $\cost$ does not enter into the one
loop corrections of $\ko$ and $\gamma$.  We recall that for hexatic
membranes the
stiffness $\gamma$ does not get renormalized\foot{This is true as
long as we are
far enough away from the vortex unbinding transition.} (just like in
the flat $XY$ model), while the stiffness $\ko$ can get stabilized at
some large value \NePe\DaGuPe. Ignoring the anisotropy, we therefore
assume that $\ko$
arrives at the fixed line $(\kappa\gamma)\eff=4\kappa\eff$.
Integrating \ebtafcn\ leads to 
\eqn\eresult{\ceff(b^{-1})=(b^{-1})^{-T/4\pi\kappa\eff}\cost
<\cost.} 
The effective chirality decreases as fluctuations are integrated out.
The result \eresult\ should not be regarded as a temperature
dependence of the effective chiral coupling since bare parameters like
$\cost$ have some 
unknown dependence on temperature.  Rather, \eresult\ gives the scale
dependence of the effective chiral coupling.  This scale dependence
may be experimentally observable, as we see in the next section.

\newsec{Tubule Radius}
The result \eresult\ has an interesting application to the tubule
structures described in Section 1.  The radius of tubules (and helical
ribbons) is given by
a competition between chirality and bending stiffness. Since chirality
tends to twist up the membrane while stiffness tends to flatten it
out, in the absence of fluctuations the radius must be given by 
$R\propto \ko/\cost$, which does have dimensions of length%
\foot{The full mean-field formula for tubule radius also involves the
$\eps^i{}_j\nabla_i m^j$ term, but its effect is similar to that of
the bulk coupling $\cost$.  We will not include this
boundary term, and so our prediction is only qualitative. Similar
remarks apply to the spontaneous edge torsion term of \Helfb.}  \HelPro.  
We can now use \eresult\ to predict how thermal fluctuations modify the
mean field theory result. We can qualitatively incorporate
fluctuations by taking the radius $R$
to be set by the effective couplings that summarize the fluctuations with
wavelengths between $\Lambda^{-1}$ and $R$ itself ($R$ supplies the
infrared cutoff stopping the running of coupling constants):
\eqn\ePNta{R\sim\kappa\eff(R)/\ceff(R)\quad.}
Experimentally, $R$ can
be quite large, \eg\ $\sim .5 \mu{\rm m}$ \Georger, while
$\Lambda^{-1} \sim 1 {\rm \AA}$.  This large scale discrepancy
$b^{-1}\sim~1000$ can enhance the effects of fluctuations in
\eresult, offsetting the smallness of $T/\ko$. 

How could one test \ePNta? In general it is not easy to adjust bare
coupling constants in a controlled way, but with $\cost$ it is easy.
To dial chirality experimentally, one can dilute the chiral molecules
with very similar but achiral analogs~\Ratpriv.  Since these molecules
have much the same shape (the chirality is very weak), we expect the non-chiral
elastic couplings describing the long scale physics of such a
system to be the same as the corresponding couplings for chiral
molecules only, while $\cost$ has decreased by the dilution fraction
$\ep$.\foot{There are some problems with this scheme. The two
molecules could phase-separate; the tubules could become unstable
after only a small amount of dilution.}
Assuming that at the scale $R$,
$\kappa\eff$ has arrived at its fixed line, \ePNta\ says that for
fixed temperature and 
fixed non-chiral elastic constants, varying $\cost$ leads to a
cylinder radius scaling law of the form 
$$R\propto (\cost)^{-(1+T/4\pi\kappa\eff)}\quad,$$
\ie\ fluctuation effects lead to an anomalous scaling law.
In terms of the dilution fraction we thus have (reinstating
Boltzmann's constant)
\eqn\eScal{R\propto\eps^{-(1+k_BT/4\pi\kappa\eff)}\quad,}
as promised in the Introduction.

In the above derivation we ignored the anisotropic terms and the total
derivative term since we do not expect those terms to qualitatively
change \eScal.  However, we must justify our omission of the
irrelevant terms:  even if we start with just the low dimension terms of
\eHshape, \eHtilt--\eHstar\ and a cutoff $\Lambda$, we expect irrelevant terms
to be generated upon integrating out all fluctuations with wavelengths
between $\Lambda^{-1}$ and $R$.  To justify this omission, we
integrate out the modes with wavelengths between $\Lambda^{-1}$ and
$(\Lambda')\inv =R/10$, say, and then argue why we can ignore the modes in the last decade.  We
start with the Hamiltonian \eHshape, \eHtilt--\eHstar, cut off at
$\Lambda^{-1}\sim 1 {\rm \AA}$.  General renormalization group ideology
assures us that at weak coupling, the effects of 
irrelevant couplings of natural
magnitude $\Order(\eps\Lambda^\Delta)$ ($\Delta$ is the canonical
dimension, $\eps$ is an overall energy scale of about one eV) can be
reproduced by retaining only the suitably adjusted relevant and
marginal couplings.  The errors incurred will be
$\Order((R\Lambda)^\Delta)$ for predictions at a scale $R$, with
$\Delta$ the canonical dimension of the least irrelevant coupling
omitted.  Now eliminate modes with wavelengths between the starting
cutoff $\Lambda^{-1}$ and some scale $\Lambda'^{-1}$ intermediate
between the cutoff and the scale $R$ of interest.  Irrelevant
couplings will be generated but with values that are $\Order(\eps
{T\over\eps}\Lambda'^\Delta)$, \ie\ suppressed from the natural values
$\Order(\ep\Lambda'^\Delta)$ by $T/\eps\sim 1/40$.  Omitting these
thermally induced irrelevant terms will induce errors of
$\Order({T\over\eps}(\Lambda'R)^\Delta)$ in predictions at the scale $R$.
If we choose $\Lambda'$ to be say $(\Lambda')\inv\sim R/10$, these errors
are small.  At this scale $\Lambda'$ the tubules are still fairly
flat and the Monge gauge calculation makes sense.  

We omit the last decade of fluctuations with wavelengths between
$\Lambda'^{-1}$ and $R$.  The irrelevant operators that would be
induced are again suppressed by $T/\eps$, while the effect on the
marginal and relevant operators of
this last decade of fluctuations is to further
renormalize them; \eg\ $\ceff(R)$ gets a factor
of $10^{-T/4\pi\kappa\eff}$ relative to $\ceff({R\over10})$.  This factor
simply modifies the constant of proportionality in \eScal.\foot{%
That is,
omitting the last decade is permissible since it slides with $R$ and
we want only the scaling with $R$. Meanwhile the {\it varying} number
of decades {\it retained} gives the anomalous scaling we seek.} Therefore
it is legitimate to ignore the irrelevant couplings for the purposes
of deriving the scaling law \eScal\ for the radius of tubules.  Note
again that the reason fluctuations did not drop out altogether is that
while $T/\eps$ is small $\log (R\Lambda)$ is not.
Of course, in the end this calculation is merely suggestive; the
difference between the two decades we included and the one decade we
omitted is not great.

\newsec{Conclusion}
Lipid bilayer membranes freely floating in solvent are a beautiful
real-world example of a statistical ensemble of random shapes. At room
temperature these systems are typically in a regime where shape
fluctuations are important but still weakly coupled, so that a
systematic perturbative expansion of their effects is possible. We
have given a systematic account of how to do such calculations, paying
particular attention to the choice of integration measure needed to
get covariant (\ie, correct) answers.

We have seen how {\it tilt} order is essential in order for the
underlying chirality of the constituent molecules (if any) to be
communicated to the macroscopic structures they form. Other forms of
in-plane order, for example hexatic order, can't do the job.  This
result may be hard to visualize from the point of view of packing
specific molecules, but we have seen how it emerges trivially from a
continuum elasticity theory: there simply are no relevant or marginal
terms available to carry chirality in the untilted case. In the tilted
case, on the other hand, chirality is the {\it only} relevant term, so
eventually it dominates the physics. The scale on which it dominates
the physics is much larger than the cutoff, however, because in the
systems of interest the chirality parameter is unnaturally small while
the stiffness is rather large. This explains the emergence of a long
scale $R$, the radius of helices and tubules.

The continuum theory also makes it straightforward to compute the
effects of thermal fluctuations on the elastic couplings. We have seen
that the effective bulk chiral coupling decreases logarithmically in
the infrared, leading to an anomalous scaling relation between
chirality and the radius $R$.

Various interesting problems remain. For example, the techniques in this paper can be used to study the effect
of fluctuations on the chiral asymmetric rippled phases studied in
\LuMacb.  Another
example comes again from tubule theory. As we approach an in-plane
phase transition we can no longer restrict our attention to the
elastic modes. The obvious place to look is at the chain-melting
transition: analogous to the smectic-C-to-A transition, we expect the
polar angle of tilt to become a new soft mode. In \Sel\ a mean-field
anomalous exponent was found describing how the tubule radius $R$
blows up at this transition. The techniques developed in this paper
should make it easy to find fluctuation corrections to this exponent.

\appendix{A}{Path Integral Measure}
The discussion in this section is
technical, and most of it matters only when we try to carry calculations
beyond one-loop order. The reader not interested in such niceties will
probably wish to skip it altogether.

To perform statistical sums we need to know how to count each configuration
just once. Our guiding principles will be that any measure $[\dd\bar\mu]$
defined on the space of geometrical data of Section~3.1, invariant under
the physical symmetries of Euclidean motions and local, will be adequate.
Any other $[\dd\bar\mu']$ will be related to $[\dd\bar\mu]$ by an invariant,
local functional of the fields; this discrepancy can be absorbed into the
Hamiltonian $H$, itself the most general invariant local functional of fields.
Accordingly let us seek one nice choice of $[\dd\bar\mu]$.

In principle this is not so difficult. We need to begin with a notion of
{\it distance} on the function space of geometrical data; given
$(\vec x (\xi),\hat m(\xi))$ and a nearby $(\vec x+\delta\vec x,\hat m+
\delta\hat m)$ we want to specify invariantly how different they are. From
this {\it metric} on field space there then follows a {\it measure}
$[\dd\bar\mu]$; if the metric is local and invariant so will be the measure.

There are also two main subtleties to this procedure. First, the passage from
metric to measure is in general complicated. Neglecting $\hat m$ for a moment,
the only invariant length we can ascribe to $\delta\vec x$ is
\eqn\emetx{\|\delta\vec x\|^2_{\vec x}\equiv
\int\dd^2\xi\sqrt{g}(\delta\vec x)^2}
where $g_{ij}=\partial_i\vec x\cdot\partial_j\vec x$. Eqn. \emetx\ is a good
starting point since if we shift, rotate, or reparameterize $\vec x$ and
$\vec x+\delta\vec x$ the length is unchanged. In fact, since $g_{ij}$
converts coordinate-space intervals into 3-space distances, \emetx\
accomplishes the objective of measuring surface fluctuations using 3-space
distance, similarly to our cutoff philosophy in Section~2.3. 

Since
$\|\cdot\|^2_{\vec x}$ depends on $\vec x$ itself, however, this metric
defines a complicated, nonlinear measure. (In contrast, an ordinary scalar
field on curved space involves
\eqn\emeth{\|\delta\phi\|^2=\int\dd^2\xi\sqrt{{\bar g}}(\delta\phi)^2\;,}
where ${\bar g}_{ij}$, while possibly complicated, is fixed and independent
of the field $\phi$.) Indeed, the resulting measure is the modified
Liouville measure of David, Distler, and Kawai \DDK\Mavra. More precisely we
are to take $\vec x$, compute $g_{ij}$, put it in conformal gauge, and
substitute the conformal factor into the DDK measure \JPAS.\foot{At first
it seems paradoxical that this measure is related to the naive one by a
{\it nonlocal} functional of $\vec x$, generated when we cast $g_{ij}$
into conformal gauge. This can happen because \emetx\ involves not just
$\vec x$ at each point $\xi$, but also its derivatives. The resulting measure
is not ``ultralocal,'' \ie\ not the product of independent measures at each
point of space. Accordingly it will be related to the naive measure (which
{\it is} ultralocal, eqn. \emeth) by a nonlocal functional.}

Fortunately, we do not need all these details. To one loop accuracy we saw
in Section~4 that our functional integrals reduce to Gaussian, saddle-point
integrals. To evaluate these it suffices to know only the metric
\emetx. Were we to work to higher loop order we would however need the full
measure.

The second subtlety involves gauge fixing. Of course we never work directly
on a space of geometrical data; instead we need to represent a surface by
some $x^a(\xi)$ and a tilt field by some $\theta(\xi)$, and there are many
ways to do this depending on our choice of coordinates $\xi^i$ and frame
$\{e_\alpha{}^i\}$. We must make sure that our $[\dd\bar\mu]$, expressed in
one ``gauge,'' gives the same physical result as in any other.

Let us warm up with the easy part of the problem, the tilt-field measure.
As mentioned earlier, the three components of the director $\hat m^a$ are really constrained
to just one angular field $\theta(\xi)$. To define it, for each surface we
need to choose a field $\{\hat e_\alpha\}$ of orthonormal frames. We make
a specific choice in \eframe, but more generally note that a different choice
$\hat e_\alpha^\prime(\xi)=R_\alpha{}^\beta(\alpha(\xi))\hat e_\beta(\xi)$
differs by an angle $\alpha(\xi)$; the same tilt field $\hat m(\xi)$ will
then be described by an angle $\theta(\xi)$ or $\theta'(\xi)=\theta(\xi)+
\alpha(\xi)$ respectively. We need an integration measure invariant under
pointwise shifts of $\theta$\foot{In particular $\alpha(\xi)=\pi$ corresponds 
to the change $\hat m \mapsto -\hat m$ under which the measure must be 
invariant (see Section 3.2).}. 
Clearly the metric
$${\|\delta\theta\|^2_{\vec x}=\int\dd^2\xi\sqrt{g}(\delta\theta)^2}\eqno\emettheta$$
analogous to \emeth\ above enjoys this shift invariance as well as invariance
under reparameterization of $\vec x$, $\theta$, and $\delta\theta$ (in fact
it is independent of $\theta$). The associated measure is thus gauge-invariant,
so we choose it.

We now turn to the more delicate matter of gauge-fixing the $\vec x$ part
of the measure. We want to reduce the redundant measure $[\dd\vx]$ for the
three degrees of freedom $\vec x(\xi)$, induced by \emetx, to a measure
$[\dd\bar\mu]$ for the one true transverse degree of freedom. The procedure
is well known \Orlando\Polch\MFM. We need to factorize $[\dd\vx]$ into the
product of a standard measure $[\dd f]$ on coordinate transformations (the
two spurious degrees of freedom) times the rest, which we will call $[\dd
\bar\mu]$:
\eqn\efacmea{[\dd\vx]=[\dd f\,]\cdot[\dd\bar\mu]}
$[d\bar\mu]$ is our desired measure. Eqns.~\ePNmd, \efacmea\ define it
once we define 
$[\dd f\,]$. 

An infinitesimal coordinate transformation
\eqn\einff{\xi^i\mapsto f^i(\xi)=\xi^i+v^i(\xi)+\Order(v^2)}
is specified by a small tangent vector field $v$ to our surface.
We may take the distance from $f$ to the identity to be
\eqn\emetv{\|v\|^2_{\vec x}=\int\dxi\sqrt{g}\,g_{ij}v^iv^j\;,}
where $g_{ij}=\partial_i\vec x\cdot\partial_j \vec x$ as usual.
Since \emetv\ is coordinate-invariant we can use it to get an invariant
distance between {\it any} two nearby $f$'s. We take this distance function to
define the measure $[\dd f\,]$; then \efacmea\ determines $[\dd\bar\mu]$ in
principle.

So far we have not made any gauge choices, so $[\dd\bar\mu]$ is gauge-invariant
as required. To represent it, however, we choose the height function $u(\xi)$
as our true degree of freedom and write an arbitrary $\vec x(\xi)$ in terms of
$u$ and $f$ as
\eqn\echvar{\vec x(\xi)=(f^1(\xi), f^2(\xi), u(f(\xi))\;.}
Then $[\dd\bar\mu]$ in \efacmea\ will be related to $[\dd u]$ (compare
\emub, where we reinstated $\theta$):
\eqn\efacmeb{[\dd\bar\mu]\equiv\CJ[u]\cdot[\dd u]\quad.}
Eqn.~\efacmeb\ defines a Jacobian factor $\CJ$ when we take and $[\dd
u]$ to be the measure defined by the metric
$$\|\delta u\|^2_u=\int \dd^2\xi\sqrt{g}\,(\delta u)^2\;.\eqno\emetu$$
Since we are in Monge gauge we take $g_{ij}=\delta_{ij}+\partial_i u\partial_j
u$ in \emetu, and similarly in \emetx, \emetv.\foot{In a different choice of
gauge, for example normal gauge, $[\dd u]$ in \efacmeb\ will get replaced by
something different, but $\CJ$ will change in a compensating way to give the
same $[\dd\bar\mu]$.} We need to compute $\CJ$.

At first it seems hard to compute $\CJ$ because it is defined in terms of
$[\dd\vx]$, $[\dd f\,]$, and $[\dd u]$ by \efacmea, \efacmeb, and each of
these has a complicated dependence on $u$ because \emetx, \emetv, and \emetu
\ all depend on $u$ via $g_{ij}$. While this is so, the {\it relation}
between the three measures is quite easy to find, as we now recall
\Polch\MFM. 
We mentioned in Section~4.1 that a field-space metric does define a
simple measure on the tangent space of field variations $\dvx(\xi)$.
To relate the three measures $[\dd\vx]$, $[\dd f\,]$, and $[\dd u]$ at
some point $\vx(\xi)$ of configuration space it suffices to relate the
corresponding measures $[\dd\dvx]$, $[\dd \dl f\,]$, and $[\dd \dl u]$
at that point. If moreover $\vx(\xi)=(\xi^1,\xi^2, u(\xi))$ is in
Monge gauge then $f$ will be close to the identity and we can define
$[\dd\dl f\,]$ by \emetv. Then \efacmea, \efacmeb\ become
$$[\dd\dvx]_u=\CJ[u]\,[\dd\dl u]_u\,[\dd v]_u\quad.$$
Defining $[\dd\dvx]_u$ by \ePNmd, $[\dd\dl u]_u$ by \emeau, and $[\dd
v]_u$ by 
\eqn\emeav{1=\int[\dd v]_u\,\ex{-\Lambda^4\eta ^2\|v\|^2}}
thus lets us compute \CCJ. We simply change variables in \ePNmd\ using
\echvar: $\dvx=(v^1,v^2,v^i\pa_iu+\dl u)$. Thus
$$1=\CJ[u]\int[\dd\dl u][\dd v]\exp\left[-\Lambda^4\eta ^2
\int \dd^2\xi\sqrt{g_0}\,(v,\delta u)T^t\,
T\left(v\atop\delta u\right)\right] \quad,$$
where $T\left(v\atop \dl u\right)=\left(v\atop \dl u+v^i\partial_iu_0\right)$.
Since $T$ is a lower-triangular matrix equal to the identity along the diagonal,
its determinant equals unity and we may drop it.
Comparing \emeau, \emeav\ we find
$$\CJ[u]\inv=\int[\dd
v]\ex{-\Lambda^4\eta ^2\int\gdxi\,v^iv^j\dl_{ij}}=\det^{-1/2}\CO_1\quad,$$
where $\CO_1$ is the multiplication operator by $\det\,g^{ij}=g\inv$.
In the language of \ePNmc\ we therefore have
\eqn\eJfinal{\CJ[u]=\prod_\xi\,{g(\xi)}^{-1/2}\quad,}
the regularized product over the points of our surface of
the volume function. 
Expression \eJfinal is not covariant but neither is Monge gauge; in
fact \CCJ\ is needed to {\it achieve} covariance of $H\eff$ as we'll
see in Appendix B.

Eqn. \eJfinal\ is our desired conversion factor; it lets us compute correct
(gauge-invariant) results using the convenient measure $[\dd u]$. We can
now make two observations needed in Section~4. Since $\CJ$ is purely
geometrical in origin, it contains no factors of the stiffness parameter
$\ko$, as claimed. Since the configuration $u(\xi)$ enters \ePNmd, \emeau,
\emeav\ only via the induced metric $g_{ij}$, $\CJ$ too depends only on
$g_{ij}$ and not on the extrinsic curvature $K_{ij}$, as we see
explicitly in eqn.~\eJfinal.  As mentioned at the end of Section 4.2
these facts imply that $\CJ$ does not enter the one loop calculation
of the renormalization of quantities like $\ko$, $\cost$ involving
$K_{ij}$.  On the other hand the correct $\CJ$ is needed to get the
right area coefficient $\mu\eff$ as we see in Appendix B below.

\appendix{B}{The Area Term}

Here we calculate the effective area coefficient $\mu\eff$ to one loop
accuracy in Monge gauge. $\mu\eff$ is a power series in the
dimensionless quantities $T/\ko$ and
$\bar\mu_0\equiv\mu_0/\Lambda^2T$; we will compute only the most
divergent part of the term of order zero in $\mob$. Our aim is to
complement the discussion in the text by showing how the constant $\eta $
appearing in the measure (eqns.~\ePNmd, \emeau, \emeav) and the
Jacobian \CCJ\ of Appendix~A enter $\meff$, and to illustrate the
general result of Section~5 that no field renormalization of the
height is needed. In particular \CCJ\ is crucial to get a result
agreeing with other gauges.

We will neglect tilt in this discussion; the changes required to
incorporate it are easy. Accordingly consider \eHshape\ with $\mo=0$.
The one-loop partition function $Z[\ub]$ comes from the saddle-point
integral 
$$Z[\ub]=\int\ddu\CJ[\ub]\,\exp\left[-{\ko\Lambda^4\over 2T}\bigl(\dl
u,\CO_\ub \dl u)_0\right]\quad,$$
where $(f,g)_0\equiv\int\dgbxi\,fg$ and \elong\ gives
\eqn\ePNba{\CO_\ub=\Lambda^{-4}\left(\Delta^2+\lfr32(\Delta u)^2\Delta
-2(\pa_i\ub)^2\Delta^2\right) +\Order(\ub^4)\quad.}
We have simplified \elong\ using rotation invariance and parts
integration. The second term of \ePNba \ gives rise to the famous
renormalization of the stiffness $\ko$; we accordingly drop it. We
also drop in the sequel terms of $\Order(\ub^4)$. We
need to compute $-T\log Z[\ub]=-T\log\CJ[\ub]+ \lfr T2\Tr\log\coub$.

Let us briefly recall how these functional traces are defined. An
operator $\CO$ has a kernel defined by $(\CO f)(\xi)=\int_{\xi'}
K_\xi(\xi')f(\xi')$, where $K_\xi(\xi')$ is a two-form at $\xi'$. Then $\Tr
\CO= \int_\xi K_\xi(\xi)$. Our operator has kernel
$K_\xi(\xi')=\Lambda^{-4}\Delta^2(1-2(\pa_i\ub)^2)\dl_\xi(\xi')$,
where the regularized covariant delta-function is
\eqn\ePNbb{\dl_\xi(\xi')=
\sum_{\la<\Lambda}\phi_\la(\xi)\phi_\la(\xi') \sqrt{\bar g(\xi')}\dxi'
\quad. }
Here $\{\phi_\la\}$ are a set of normalized eigenvectors for the
covariant Laplacian, $\Delta_{\bar g}\phi_\la=-\la^2\phi_{\la}$ with
$\|\phi_\la\| ^2_{\bar g}=1$. In \ePNbb\ we have regularized by
cutting off $\la$; to integrate out a shell in momentum space we
actually want only the contribution of modes between $\Lambda$ and
$b\Lambda<\Lambda$. 

In flat space one has $\phi_\la(\xi)\propto\sin k_i\xi^i$ or $\cos
k_i\xi^i$, $\la^2=k_ik_jg^{ij}=g^{-1/2}k_ik_i$ in coordinates where
$g_{ij} =\sqrt g\dl_{ij}$ is constant. For {\it nearly} flat space
$\phi_\la$ and $\la$ will get corrections, but for $\Lambda$ near the
cutoff they are suppressed by 
powers of (curvature/$\Lambda$) and we are computing only the leading
divergence. Accordingly, our determinant becomes 
$$\lfr T2\Tr\log\coub = \lfr T2\int\dgbxi
\int_{b\Lambda\gb^{1/4}}^{\Lambda\gb^{1/4}}\dk\left[
\log\lfr{k^4}{\Lambda^4}+\log \bigl(1-2(\pa \ub)^2\bigr)\right]\quad.$$
For a thin shell of width $1-b=\ep$, keeping only terms linear in
$\ep$ (and still up to $\Order(\ub^2)$) we find
$$\lfr T2\Tr\log\coub=\lfr T2\int\dxi\int_{b\Lambda}^\Lambda\dk
\left[\log\gb- 2(\pa\ub)^2\right]\quad.$$
Applying the same analysis to the operator $\CO_1$ in Appendix A we find
\eqn\ePNbc{\eqalign{
-T\log Z[\ub]&=\lfr T2\int\dxi\int_{b\Lambda}^\Lambda\dk
\left[-2(-\half)(\pa\ub)^2+(\pa\ub)^2-2(\pa\ub)^2\right]\cr
&=0\quad.\cr}}
This is the same answer as found by David and Leibler in normal
gauge~\DaLe, starting from the same measure \ePNmd.

We cannot yet conclude however that the quadratically-divergent bit of
the renormalization of the area term in the effective Hamiltonian
$H\eff$ vanishes. Having integrated out a shell in momentum space we
have found an $H'[u]$ which reproduces the answers of $H[u]$ when we
thin out the degrees of freedom in \ePNmc. This is not quite the same
thing as changing the cutoff, however, because $\Lambda$ appears
explicitly in \ePNmc\ as well as in the spacing of points $\xi$. The
difference between $[\dd\dvx]_{b\Lambda}$ and the thinned-out measure
$[\dd\dvx] _\Lambda^{\rm slow}$ is the product over all points of $b\inv$,
or 
\eqn\ePNbd{H\eff=H'-T\left({\Lambda\over\pi}\right)^2
\log b\inv\int\dgbxi\quad.}
Thus while we found $H'$ has no area term, $H\eff$ shows that 
$$\meff=-T\ep{\Lambda^2\over\pi^2}$$
plus less divergent terms. With a different choice of the arbitrary
constant $\eta $ in \ePNmd\ we could of course manage to cancel even this
contribution. 

Generalizing to nonzero bare $\mo$, we find by dimensional analysis no
new quadratically-divergent terms:
$$\meff=\mo-T\ep{\Lambda^2\over\pi^2}\quad,$$
plus less-divergent terms, as claimed at the end of Section 5.

We remark that \ePNbd\ is covariant as it must be. Had we omitted
\CCJ\ from \ePNbc\ this covariance would have been spoiled. One could
imagine trying to cure the problem by rescaling the field $u$, but the
Ward identity of Section~5.2 shows that this rather desperate gambit
will not work. Indeed we can now do perturbation theory in $\mo$ to
work out $\meff$ in powers of the dimensionless parameter $\mob=\mu_0/
\Lambda^2T$ (still to one-loop order in
$T/\ko$); to all orders the answer turns out to be covariant with no
field rescaling~\ref\CNP{W. Cai, P. Nelson, and T. Powers, unpublished (1993).}.

\vskip1truein \leftline{\bf Acknowledgements}
{\frenchspacing\noindent
We would like to thank R. Bruinsma,
F. David, M. Goulian, E. Guitter,
J. Distler,
F. MacKintosh,
S. Milner,
D.R. Nelson,
J. Polchinski,
J. Preskill,
B. Ratna,
C. Safinya,
J. Schnur,
J. Selinger, 
R. Shashidar,
E. Wong,
and especially S. Amador, T. Lubensky, and J. Toner for innumerable
discussions.} This work was supported in part by NSF grant PHY88-57200, the
Petroleum Research Fund, and the A.~P.~Sloan Foundation. P.N.~thanks
the Aspen Center for Physics and the University of California, Santa Barbara
(NSF grant PHY-9157463) for hospitality during the course of this work.

\footatend\vfill\supereject\immediate\closeout\rfile\writestoppt
\baselineskip=14pt\centerline{{\bf References}}\bigskip{\frenchspacing%
\parindent=20pt\escapechar=` \input refs.tmp\vfill\eject}\nonfrenchspacing
\bye